\documentclass[aps,prl,twocolumn,showpacs,superscriptaddress,groupedaddress,floatfix]{revtex4-1}

\usepackage{graphicx,psfrag}
\usepackage{mathrsfs}
\usepackage{amsmath,amsfonts,amssymb}
\usepackage{multirow}
\usepackage{comment}
\usepackage{ulem}
\usepackage{hyperref}
\usepackage{enumitem}
\usepackage{morefloats}
\usepackage{bm}

\def\IncludeSM{1}

\newcommand{\be}{\begin{equation}}
\newcommand{\ee}{\end{equation}}
\newcommand{\bel}{\begin{align}}
\newcommand{\eel}{\end{align}}

\def\lm{\ell m}
\def\p{\partial}
\def\non{\nonumber}                     
\def\e{{\rm e}}
\def\i{{\rm i}}
\def\f{{\rm f,~BH}}
\def\gccm{{\rm g\,cm^{-3}}}
\def\Msun{{M_{\odot}}}
\def\Mpc{{\rm Mpc}}
\def\GMc2{{\rm G M_{\odot} c^{-2}}}
\def\Mpc{{\rm Mpc}}
\def\eps{\epsilon}
\def\eps{\epsilon}
\def\B{\mathcal{B}}
\def\I{\mathcal{I}}
\def\M{\mathcal{M}}
\def\O{\mathcal{O}}
\def\R{\mathcal{R}}
\def\vareps{\varepsilon}
\def\vrho{\varrho}
\def\check{$\checkmark$}
\def\cross{$\times$}
\def\l{\ell}
\def\lm{{\ell m}}
\def\hE{\hat{E}}
\def\mns{M_\text{NS}}
\def\Rns{R_\text{NS}}
\def\kt2{\kappa^\text{T}_2}
\def\Mmax{M_\text{TOV}^\text{max}}
\def\Mo{{\rm M_{\odot}}}
\def\kt2{\kappa^\text{T}_2}
\def\Mmax{M_\text{max}}
\def\Rmax{R_\text{max}}
\def\Rmax{R_\text{max}}
\def\Mmaxll{M_\text{max}^\text{LB}}

\def\params{{\boldsymbol{\theta}}}
\def\F{\mathcal{F}}
\def\Fbar{\bar{\mathcal{F}}}
\def\MM{\Fbar}

\newcommand{\nrpm}{\texttt{NRPM}}
\newcommand{\TEOB}[1]{\texttt{TEOBResumS{#1}}}

\usepackage{color}
\definecolor{cyan}{rgb}{0,0.9,0.9}
\definecolor{orange}{rgb}{0.9,0.5,0}
\definecolor{magenta}{rgb}{1,0,1}
\definecolor{purple}{rgb}{0.8,0.4,0.8}
\definecolor{gray}{rgb}{0.5,0.5,0.5}
\newcommand{\mb}[1]{{\textcolor{blue}{\texttt{MB: #1}} }}
\newcommand{\bs}[1]{{\textcolor{green}{\texttt{SB: #1}} }}
\newcommand{\rg}[1]{{\textcolor{orange}{\texttt{RG: #1}} }}
\newcommand{\an}[1]{{\textcolor{red}{\texttt{AN: #1}} }}

\newcommand{\todo}[1]{{\textcolor{red}{TODO: [#1]}}} 
\newcommand{\red}[1]{{\textcolor{red}{#1}}} 
\newcommand{\newtxt}[1]{{\textcolor{red}{#1}}} 
\newcommand{\oldtxt}[1]{{\textcolor{gray}{\sout{#1}}}} 
\newcommand{\oldnewtxt}[2]{{\textcolor{gray}{\sout{#1}}}\red{#2}} 
\newcommand{\timesto}[1]{\times 10^{#1}}

\begin{document}

\title{Fast, faithful, frequency-domain effective-one-body
  waveforms for compact binary coalescences}

\author{Rossella \surname{Gamba}${}^{1}$}
\author{Sebastiano \surname{Bernuzzi}${}^{1}$}
\author{Alessandro \surname{Nagar}${}^{2,3,4}$}

\affiliation{${}^1$Theoretisch-Physikalisches Institut, Friedrich-Schiller-Universit{\"a}t Jena, 07743, Jena, Germany}  
\affiliation{${}^2$Centro Fermi - Museo Storico della Fisica e Centro Studi e Ricerche Enrico Fermi, Rome, Italy}  
\affiliation{${}^3$INFN Sezione di Torino, Via P. Giuria 1, 10125 Torino, Italy}  
\affiliation{${}^4$Institut des Hautes Etudes Scientifiques, 91440 Bures-sur-Yvette, France}

\date{\today}

\begin{abstract}
  The inference of binary neutron star properties from
  gravitational-wave observations requires the generation of millions
  of waveforms, each one spanning about three order of magnitudes 
  in frequency range. Thus, waveform models must be efficiently
  generated and, at the same time, be faithful from the post-Newtonian 
  quasi-adiabatic inspiral up to the merger regime.
  A simple solution to this problem is to combine effective-one-body
  waveforms with the stationary phase approximation to obtain
  frequency-domain multipolar approximants valid from any low
  frequency to merger.
  We demonstrate that effective-one-body frequency-domain waveforms 
  generated in post-adiabatic approximation are computationally competitive 
  with current phenomenological and surrogate models, (virtually) arbitrarily 
  long, and faithful up to merger for any binary parameter. 
  The same method can also be used to efficiently generate intermediate 
  mass binary black hole inspiral waveforms detectable by space-based interferometers.
\end{abstract}

\maketitle

Gravitational-wave (GW) analyses of binary neutron star (BNS) signals rely on matched filtering 
techniques and waveform models to infer the source properties from observations.
The inspiral-to-merger signal is observable in the ground-based interferometer 
frequency band for minutes, that correspond to thousands of
inspiralling cycles to merger~\cite{TheLIGOScientific:2014jea,
  TheVirgo:2014hva}. 
Waveform templates must model the signal phase evolution 
over a frequency range spanning
from few Hz to kHz~\cite{Damour:2012yf, Bernuzzi:2014owa, Breschi:2019srl}. 
Further, due to the large number ($\sim 10^7$) of waveforms needed to explore 
the posterior distribution of the parameters, such models
also need to be computationally efficient.
Similar issues  arise for the computation of waveforms for binary 
black hole (BBH) inspirals, that are observable in the mHz to Hz
and dHz regime with space-based interferometers for binary
masses ${\sim}(100-10^{5})\Mo$~\cite{Audley:2017drz,Sedda:2019uro,Kawamura:2020pcg}. 
In this case, the waveforms efficiency
requirements are even more demanding as the binary remains 
in band for days to years, corresponding to up to millions
inspiralling cycles~\cite{Sesana:2016ljz}.

Analytical post-newtonian (PN) approximants are quick to evaluate and can be turned 
into closed-form frequency-domain templates by applying the stationary 
phase approximation (SPA)~\cite{Damour:1997ub,Damour:2000gg,Damour:2000zb,Buonanno:2009zt,Blanchet:2013haa}.
While PN approximants become unfaithful as the binary motion becomes nonadiabatic (high
velocities regime), the domain of validity of the SPA itself   
was proven to be accurate at least up to frequencies corresponding to
the last stable orbit, e.g.~\cite{Damour:2000gg,Buonanno:2009zt}. 
Such a validity interval corresponds to binaries with total mass $M\lesssim 13\Mo$ 
in the ground-based interferometers range, and it  is large enough to cover 
BNS and - possibly - also light BHNS or BBH systems. From the trivial scaling 
of the waveform with the binary mass, it follows that the validity
of the SPA holds also for the inspiral of stellar and 
intermediate-mass BBH~\cite{Audley:2017drz,Sedda:2019uro}. 

Beyond PN approximants, effective-one-body
(EOB)~\cite{Buonanno:1998gg,
  Buonanno:2000ef,Damour:2000we,Damour:2001tu,Damour:2008qf,Damour:2009kr,Damour:2015isa,Bohe:2016gbl,Bini:2019nra,Bini:2020wpo,Bini:2020nsb,Nagar:2020pcj,Ossokine:2020kjp}
and phenomenological (Phenom)~\cite{Hannam:2013oca,Khan:2015jqa,London:2017bcn,Garcia-Quiros:2020qpx,Pratten:2020fqn,Pratten:2020ceb} models describe the waveform from the early inspiral up to 
merger [and ringdown, if dealing with BBH systems], but are in general 
computationally less efficient. While Phenom models output 
frequency domain (FD) waveforms, EOB models natively generate time domain
(TD) waveforms, so that an additional Fourier transform is needed, with
the related performance loss.
Reduced-order modeling techniques offer a solution to the issue of
performances as they can be used to produce fast surrogate models from
a training set of waveforms~\cite{Field:2011mf,Purrer:2015tud,Lackey:2016krb,Lackey:2018zvw,Cotesta:2020qhw}. 
However, surrogate waveforms are limited by the length and parameters
span of the training set and they must be regenerated if
the baseline model is varied.

EOB waveforms can alternatively be speeded up using dedicated
analytical methods. The postadiabatic method gives an approximate,
iterative solution for the EOB (circular) Hamiltonian
dynamics~\cite{Nagar:2018gnk,Akcay:2018yyh,Nagar:2018plt}. Such a
simple, physically motivated technique was shown to allow for 
the computation of BNS TD waveforms  
from frequencies as low as a few Hz in a matter of tens of milliseconds. 
However, such waveforms still need to be translated in the FD.
In this work we apply the SPA to EOB waveforms in order to obtain 
computationally inexpensive frequency-domain templates. The FD
waveforms obtained this way are suitable for the GW data analysis of
BNS signal up to merger and of long BBH inspiral for masses ${\gtrsim}1000\Mo$.
Throughout this work $M$ is the binary mass, $q\geq1$ the
mass ratio, $m_i$ ($i=1,2$) the individual masses, $\nu=m_1m_2/M^2$, $\chi_i\equiv S_i/m_i^2$ the
dimensionless individual spins (anti)aligned with the orbital angular momentum, 
$\Lambda_i$ the individual quadrupolar tidal polarizability 
parameters~\cite{Damour:1983a,Hinderer:2007mb,Damour:2009vw,Binnington:2009bb}, 
and $\tilde\Lambda$ the reduced 
tidal parameter~\cite{Flanagan:2007ix,Damour:2009wj,Favata:2013rwa}. 
Geometric units $G=c=1$ are employed unless stated differently.

The FD extension of a given a TD EOB waveform is computed by applying the 
SPA to the multipolar TD modes $h_{\ell m}(t)=a_{\lm} e^{\i\phi_\lm(t)}$ to obtain
\be 
\label{eq:spa_modes}
\tilde{h}_{\ell m}^{\rm SPA} = \tilde{A}^{\rm SPA}_{\ell m} e^{i
  \Psi^{\rm SPA}_{\ell m}} =
\frac{a_{\lm}\left(t_{f}\right)}{\sqrt{\ddot{\phi}_{\ell
      m}\left(t_{f}\right)/2\pi}} e^{{\rm
    i}\left[\psi_{f}\left(t_{f}^{\ell m} \right)-\pi / 4\right]} \,, 
\ee 
with $\psi_{f}(t) \equiv 2 \pi f t-\phi(t)$
and where $t_f$ denotes the saddle point of $\psi_f(t)$.
The two GW polarizations in FD are then computed 
by combining the multipolar modes with spin-weighted
spherical harmonics in a standard way. 
The TD-to-FD computation is straightforward, the key technical
details in our implementations are given in the Supplemental Material.
We apply the SPA to TD \TEOB{}, a state-of-the-art effective EOB approximant for 
spin-aligned
binaries~\cite{Damour:2014sva,Nagar:2017jdw,Nagar:2018zoe,Nagar:2018plt,Nagar:2019wds,Nagar:2020pcj},
that includes resummed tidal interactions from PN and gravitational-self force 
results~\cite{Bernuzzi:2015rla,Nagar:2018plt,Akcay:2018yyh}.
We shall show that the FD \TEOB{PA} retains the same accuracy as the TD \TEOB{} 
up to merger for any BNS signal, and that for waveforms with initial frequency 
$f_0\lesssim15$~Hz its speed is comparable to ${\tt SEOBNRv4Tsurrogate}$. 

\begin{figure*}[t]
  \centering 
  \includegraphics[width=0.49\textwidth]{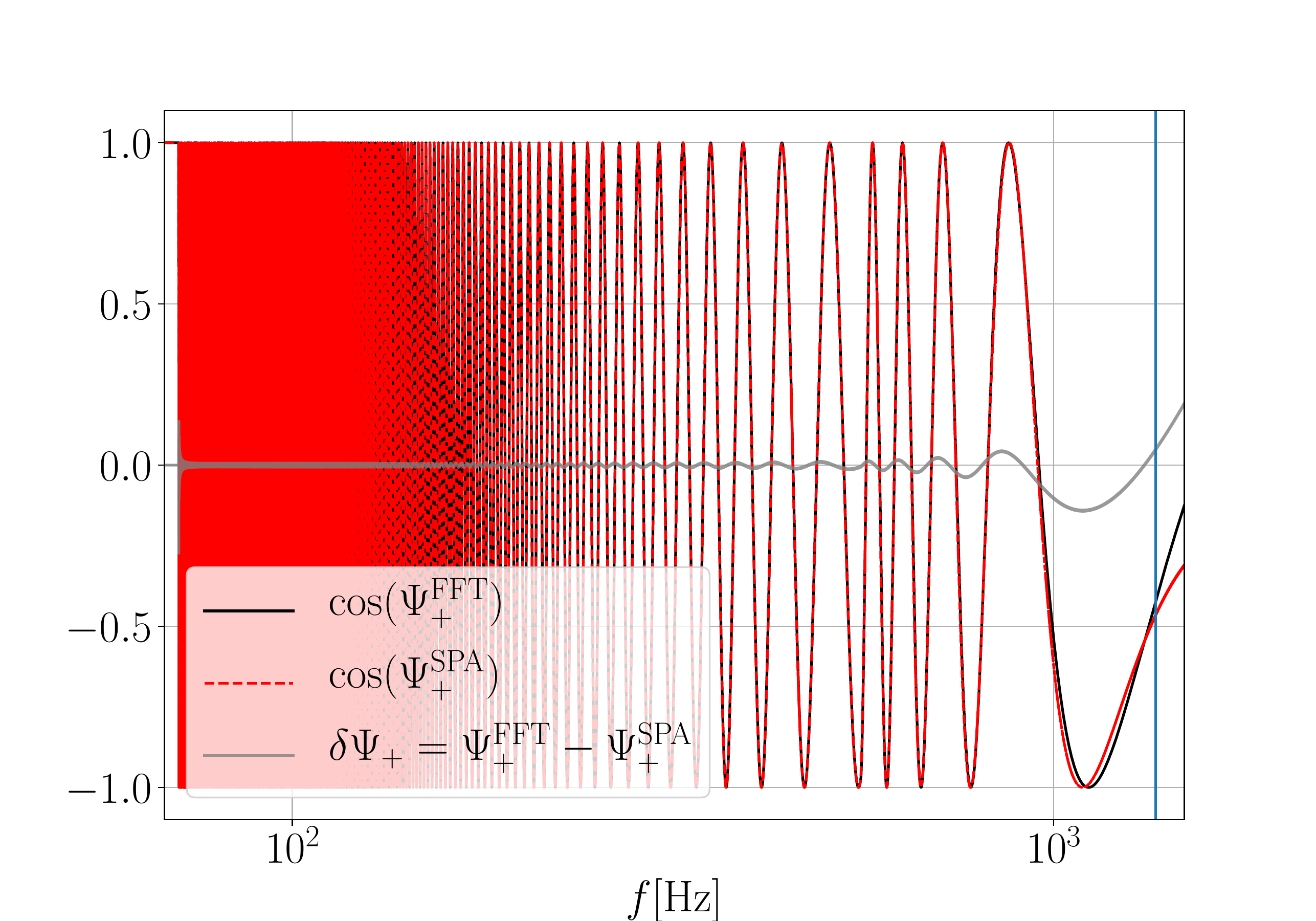}
  \includegraphics[width=0.49\textwidth]{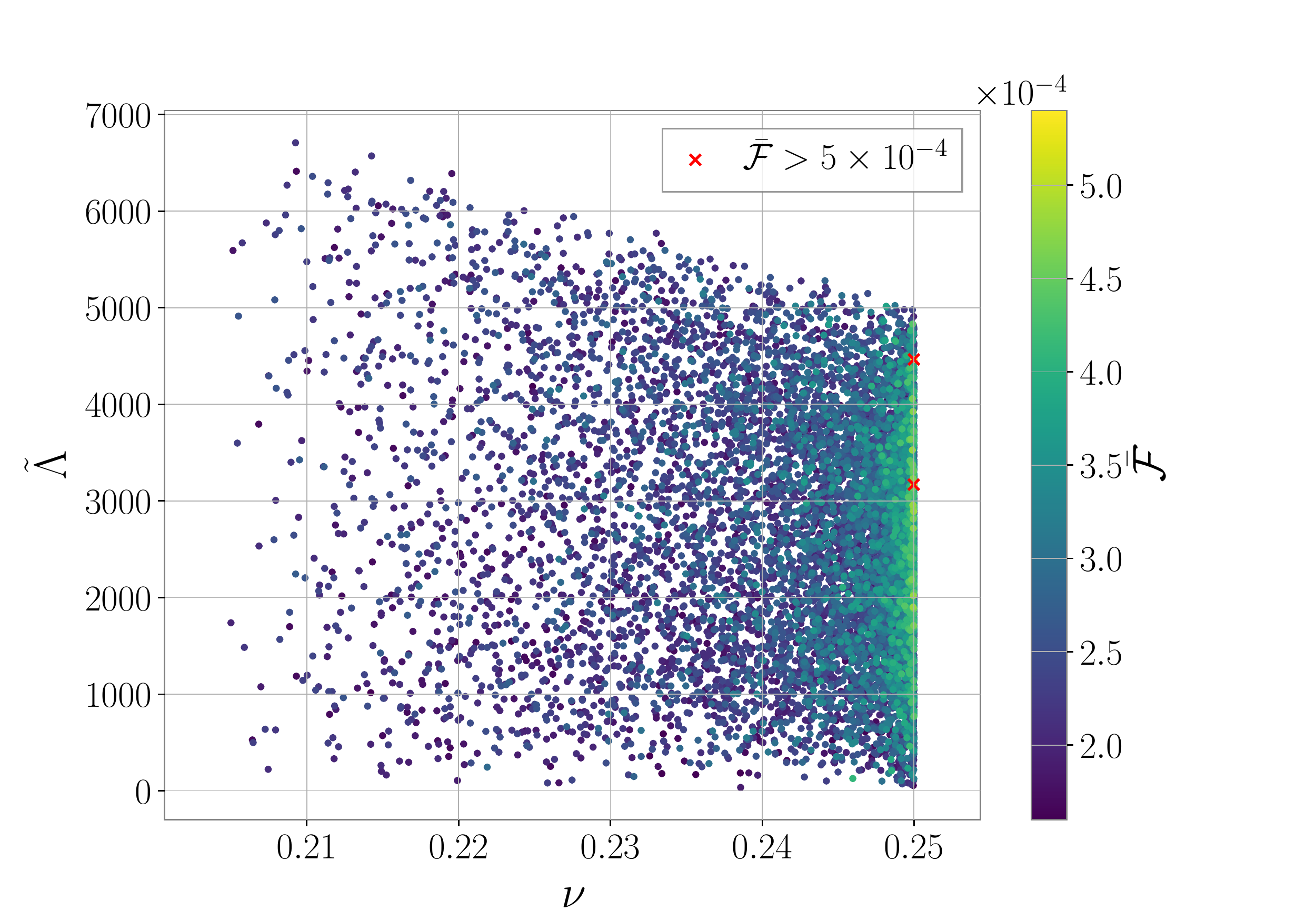}
  \caption{Left: comparison between the (cosines of) 
    the frequency domain phase of the $\tilde{h}_+$, computed with {\tt TEOBResumSPA} 
    and the FFT of {\tt TEOBResumS} for a $q=1$, $M=2.8 \Msun$, $\chi_{1,2}
    = 0.1$ and $\tilde{\Lambda} = 400$ system. The phase difference
    between the two models remains below $\delta\Psi_+\lesssim
    0.1\,$rad to merger (blue vertical line).
    Right: Unfaithfulness between {\tt TEOBResumSPA} and Fourier-transformed {\tt TEOBResumS}, 
    computed for $10^4$ binary systems with varying spins, 
    masses and component tidal parameters. Waveforms are computed from an initial frequency of $20$~Hz, 
    and matches computed between $20$~Hz and $2$~kHz. 
    We find that only two values (denoted with red crosses) lie below
    the SNR $80$ threshold of Eq.\eqref{eq:threshold}. The mismatch values found, 
    however, are still $\bar\F\sim10^{-4})$.}
  \label{fig:BNS_ex}
\end{figure*}

Figure~\ref{fig:BNS_ex} (left panel) shows the FD phasing of a fiducial BNS waveform for a $(M/M_\odot, q, \chi_1, \chi_2, \tilde{\Lambda}) = (2.8,1,0,0, 400)$ 
system, sampled at $8192$~Hz with initial frequency $f_0 = 20$~Hz and computed either Fourier transforming 
the TD \TEOB{} (solid black line) or using the SPA (dashed red line). 
\TEOB{PA} correctly reproduces the waveform of the original TD model from
the early inspiral up to merger; the accumulated total
phase difference amounts to ${\sim}0.1$ to merger (vertical line).

We quantitatively assess the faithfulness of \TEOB{PA} against 
\TEOB{} on a sample of $10^4$ waveforms from BNS with $m_{1,2} \in [1, 2.5] M_{\odot}$, 
$\chi_{1,2} \in [-0.5, +0.5]$ and $\Lambda_{1,2} \in [10, 5000]$.
Since the masses and the tidal
parameters are sampled separately, no specific equation of state is imposed.
We recall that the mismatch $\bar\F$ (and the match $\F$) between 
two waveforms $(h,s)$ is defined by
\be
\label{eq:barF}
\bar\F \equiv  1-\F=1- \max_{t_c, \phi_c} \frac{(h, s)}{\sqrt{(h,h)(s,s)}} \ ,
\ee
where $t_c$ and $\phi_c$ denote the time and phase at coalescence, 
and the Wiener scalar product
associated to the power-spectral density (PSD) of the
detector, $S_n(f)$, is
\be
(h, s) = 4\ \Re{ \int \! \frac{\tilde{h}^*(f) \tilde{s}(f)}{S_n(f)} \, \mathrm{d}f.}
\ee
Since the detection rate loss scales as $(1 - (1 - \bar\F)^3)$, 
$\bar\F\leq 0.035$ is usually usually regarded to be satisfactory for detection
purposes~\cite{Lindblom:2008cm}. However, the value of $\bar{\F}$ does not depend 
on the signal SNR and does not account properly for statistical fluctuations (or
lack thereof) due to the background noise. Two waveforms are faithful
if~\cite{Lindblom:2008cm,Damour:2010zb, Chatziioannou:2017tdw}
\be
\label{eq:threshold}
\bar{\F} \leq \bar{\F}_{\rm SNR} \equiv \frac{D}{2\, \mathrm{SNR}^{2}} \ ,
\ee
where $D=6$ is the number of intrinsic parameters.
This means that we require for the systematical errors introduced by
the SPA approximation to be smaller than the expected statistical fluctuations.
The threshold SNRs chosen in Eq.~\eqref{eq:threshold} are 
$13$, $33$ and $80$.
The first two values mimic the SNRs
of the two BNSs observed by LIGO/Virgo in O3 and O2 respectively,
while the last value can be reached for GW170817-like event at design
sensitivity or in third generation detectors~\cite{Gamba:2020wgg}. 
The above numbers lead to the threshold unfaithfulnesses of
$\bar{\F}_{13} \approx  1.8 \times 10^{-2}$, 
$\bar{\F}_{33} \approx 2.7 \times 10^{-3} $ and $\bar{\F}_{80} \approx 5\times 10^{-4}$.

Figure~\ref{fig:BNS_ex} (right panel) shows the matches between \TEOB{} and \TEOB{PA}
for our BNS sample with a starting frequency $f_0 = 20$~Hz and assuming
the Advanced LIGO {\tt ZeroDetunedHighPower} PSD~\cite{Sn:advLIGO}.
We find that more than $99\%$ of total mismatches lie below the most conservative threshold $\bar{\F}_{80}$. 
The worst performances ($\bar{\mathcal{F}}\sim 5.2 \times 10^{-4}$ and $\bar{\mathcal{F}} \sim 5.4 \times 10^{-4}$) 
are obtained for two cases with equal mass configurations and large
$\tilde{\Lambda}$ values ($\tilde{\Lambda} > 2000$).

While the loss of accuracy with respect to the
TD EOB is negligible, the speed up given by the SPA is significant.
The computational performance of
our FD EOB waveform is assessed by comparing the evaluation 
times of \TEOB{PA} waveforms to  that of other FD BNS approximants.
In particular, we compare to PN {\tt TaylorF2}, considered here with 
a 3.5PN-accurate description of the point-mass phasing and 7PN-accurate tidal effects; 
${\tt IMRPhenomDNRTidal}$~\cite{Husa:2015iqa, Dietrich:2017aum}, a phenomenological approximant 
for aligned-spin binaries augmented by the {\tt NRTidal} phase and amplitude prescriptions; 
{\tt SEOBNRv4Tsurrogate}~\cite{Lackey:2018zvw}, 
a FD surrogate of the EOB tidal model of Ref.~\cite{Hinderer:2016eia, Steinhoff:2016rfi}; ${\tt SEOBNRv4\_ROM\_NRTidal}$ 
a FD reduced-order model of the EOB model \cite{Bohe:2016gbl} augmented by the ${\tt NRTidal}$ phasing.

\begin{figure}[t]
  \centering 
  \includegraphics[width=0.45\textwidth]{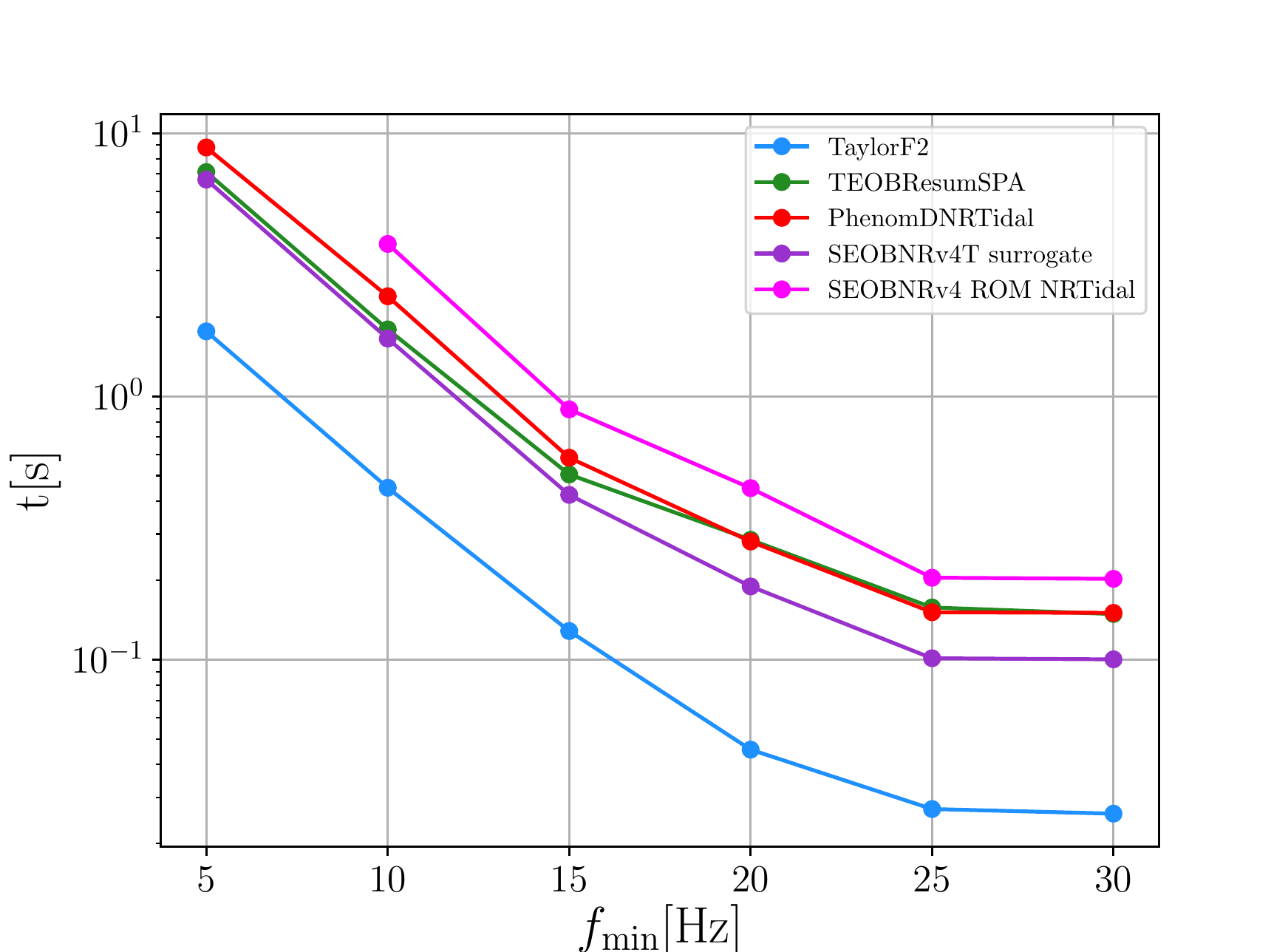}
  \caption{Generation times of different BNS waveforms averaged over 10 repetitions as a
    function of initial frequency for a fiducial BNS with $(M/\Msun, q,
    \chi_1, \chi_2, \tilde{\Lambda}) = (2.8, 1, 0, 0, 400)$. 
    Below $15$ Hz {\tt TEOBResumSPA} is comparable to the {\tt SEOBNRv4Tsurrogate} and
    {\tt IMRPhenomDNRTidal} models and ${\sim}2$
    times faster than {\tt SEOBNRv4\_ROM\_NRTidal}.}
  \label{fig:timing}
\end{figure}

We simulate the fiducial BNS 
coalescence with varying initial frequency 
$f_0= (5, 10, 15, 20, 25, 30)$~Hz and uniform 
frequency spacing $\Delta f = (1/8192, 1/2048, 1/512, 1/256, 1/128, 1/128)$~Hz,
and compute the average generation time to $f_{\rm max} = 2048$~Hz 
over ten repetitions for each of the approximants listed above.  
The timing results are shown in Fig.~\ref{fig:timing}.
\TEOB{PA} generation times for $f_0\gtrsim15$~Hz, in green, are
comparable to those of {\tt IMRPhenomDNRTidal}, 
a factor two smaller than {\tt SEOBNRv4\_ROM\_NRTidal} and up to a factor ten slower than {\tt TaylorF2}.
For starting frequencies  $f_0<15$~Hz, \TEOB{PA} is comparable to {\tt SEOBNRv4Tsurrogate} and {\tt IMRPhenomDNRTidal},
and at least a factor two faster than {\tt SEOBNRv4\_ROM\_NRTidal}.

Further tests indicate that \TEOB{PA} generation times display a small variance over 
a large parameter space ($m_{1,2} \in [1, 2.5] \Msun, \chi_{1,2} \in [-0.5, 0.5], \Lambda_{1,2}\leq 5000$). 
This is due to the fact that the largest 
computational burden of the approximant lies in the uniform-frequency 
interpolation of the multipoles $h_{\ell m}$, which is approximately independent 
of the binary parameters. A further speedup would be simply obtained by considering 
a non-uniform frequency grid for the likelihood computation, see e.g.~\cite{Zackay:2018qdy, Vinciguerra:2017ngf}. 
We also stress that, being the SPA analytical, the waveform is valid
for any initial frequency and any binary parameters. This is in contrast
to waveform models based on surrogates or machine learning, that are limited 
by the length and parameter range of the training set.
Note for example, that it is not possible to generate 
{\tt SEOBNRv4\_ROM\_NRTidal} waveform at dimensionless frequencies below $Mf=9.85\times 10^{-5}$,
which is the minimal frequency of the surrogate.
Thus, the flexibility of the SPA method applied to EOB waveforms has an
obvious advantage, in particular in view of the continuous and rapid development of EOB models. 

\begin{table}[t]
  \centering    
  \caption{Source properties of GW170817 \cite{Abbott:2018wiz} and
    GW190425 \cite{Abbott:2020uma}. The estimates found with \TEOB{PA}
    are compared to the LIGO-Virgo results. The \TEOB{PA} anaysis is
    performed up to $\sim 1$kHz following 
    \cite{Gamba:2020wgg} in order to minimize waveform systematics;
    this approach (not the waveform model) is responsible for the
    slightly different estimate of $\tilde{\Lambda}$ when compared to \cite{Abbott:2018wiz}.}
\scalebox{0.9}{
  \begin{tabular}{c|cc|cc}        
    \hline\hline
    & \multicolumn{2}{c|}{GW170817} & \multicolumn{2}{c}{GW190425} \\
    & LVC \cite{Abbott:2018wiz} & {\tt TEOBResumSPA} & LVC \cite{Abbott:2020uma} & {\tt TEOBResumSPA} \\
    \hline
     $1/q$ & 0.73-1.00 & $0.89^{+0.10}_{-0.2}$ 
    & 0.8-1.0 & $0.9^{+ 0.1}_{-0.1}$ \\
    $\mathcal{M}$ [$\Msun$]& $1.1975^{+0.0001}_{-0.0001}$ & $1.1976^{+0.0001}_{-0.0001}$ & $1.4868^{+0.0003}_{-0.0003}$ & $1.4868^{+0.0005}_{-0.0005}$\\
    $\tilde{\Lambda}$\footnote{The values of $\tilde{\Lambda}$ are quoted after reweighting to flat in $\tilde{\Lambda}$ prior. 
      Note however that the procedures followed for the reweighting of the GW170817 and GW190425 posteriors are different, for consistency with Sec. 3D of \cite{Abbott:2018wiz} 
      and App. F.1 of \cite{Abbott:2020uma}.} & $300^{+500}_{-190}$ & $530^{+350}_{-310}$ &$\leq 600$ & $\leq 550$\\ 
    $\chi_{\rm eff}$ & $0.00^{+0.02}_{-0.01}$ & $0.00^{+0.02}_{-0.01}$ & $0.01^{+0.01}_{-0.01}$& $0.01^{+0.01}_{-0.01}$\\
    $D_L$ [Mpc]& $39^{+7}_{-14}$ & $42^{+6}_{-13}$& $159^{+69}_{-72}$ & $180^{+61}_{-77}$\\
    \hline\hline
  \end{tabular}
}
\label{tab:PE}
\end{table}

We demonstrate the use of \TEOB{PA} in GW parameter
estimation by performing the analysis of
GW170817~\cite{TheLIGOScientific:2017qsa,GBM:2017lvd, Abbott:2018wiz, LIGOScientific:2018mvr} and
GW190425~\cite{Abbott:2020uma}. 
We employ the {\tt pbilby}~\cite{Smith:2019ucc, Romero-Shaw:2020owr} and
{\tt dynesty}~\cite{2020MNRAS.493.3132S} parameter estimation
infrastructures and the same setup of~\cite{Gamba:2020wgg}, 
to which we refer for all technical details.
These analyses run in two-to-three days time on 4 $\times$ 16 CPUs.
For comparison, the same GW170817 analysis, performed with the {\tt
IMRPhenomPv2NRTidal} \cite{Khan:2015jqa, Dietrich:2017aum} approximant with an identical setup, 
required longer than 5 days to complete. Results are listed 
in Table~\ref{tab:PE}, and are consistent with LIGO-Virgo analyses. 
The measurement of $\tilde\Lambda$ differs from the LIGO-Virgo 
one because of our conservative choice of the sampling rate, 
aimed at minimizing high-frequency systematic effects 
as discussed in~\cite{Gamba:2020wgg}. 

Fast and accurate EOB waveforms will be crucial for the analysis of
high-SNR BNS signals as those detectable by the Einstein Telescope or
Cosmic Explorer \cite{Punturo:2010zza, Sathyaprakash:2011bh, Maggiore:2019uih, Reitze:2019iox}. Recent work has demonstrated that at SNR
${\gtrsim}80$ the systematics among advanced BNS approximants shown in
Fig.~\ref{fig:timing} will be the 
dominant source of error in the measurements of tidal effects \cite{Gamba:2020wgg}.
High-precision measurements for constraining the NS equation of state will
require new numerical-relativity informed
tidal EOB models \cite{Bernuzzi:2014owa,Hinderer:2016eia} or new closed-form representations of the
tidal sector~\cite{Dietrich:2017aum}. Ongoing work in these directions based on
the EOB SPA method will be presented elsewhere.

\begin{figure}[t]
  \centering 
  \includegraphics[width=0.49\textwidth]{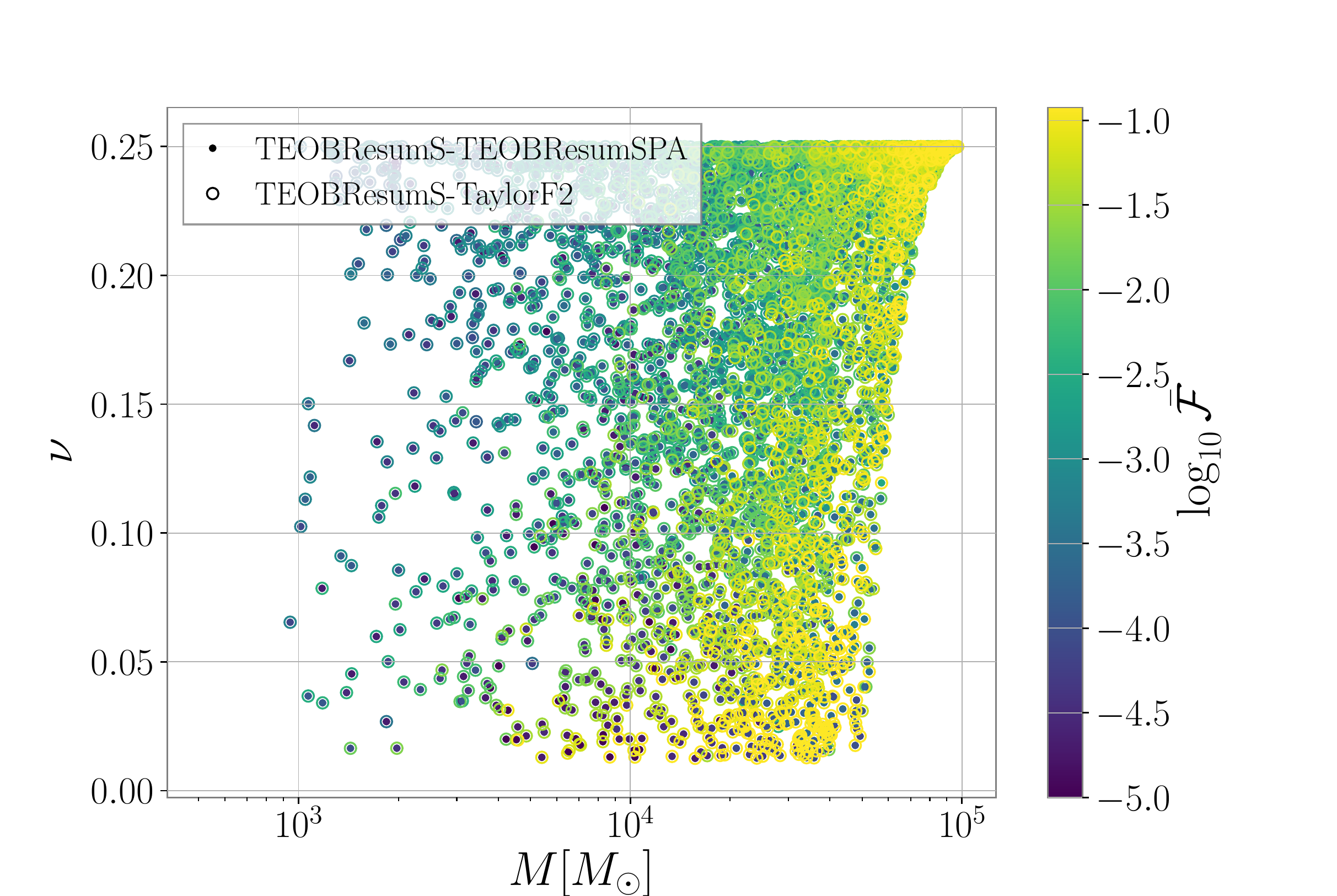}
  \includegraphics[width=0.49\textwidth]{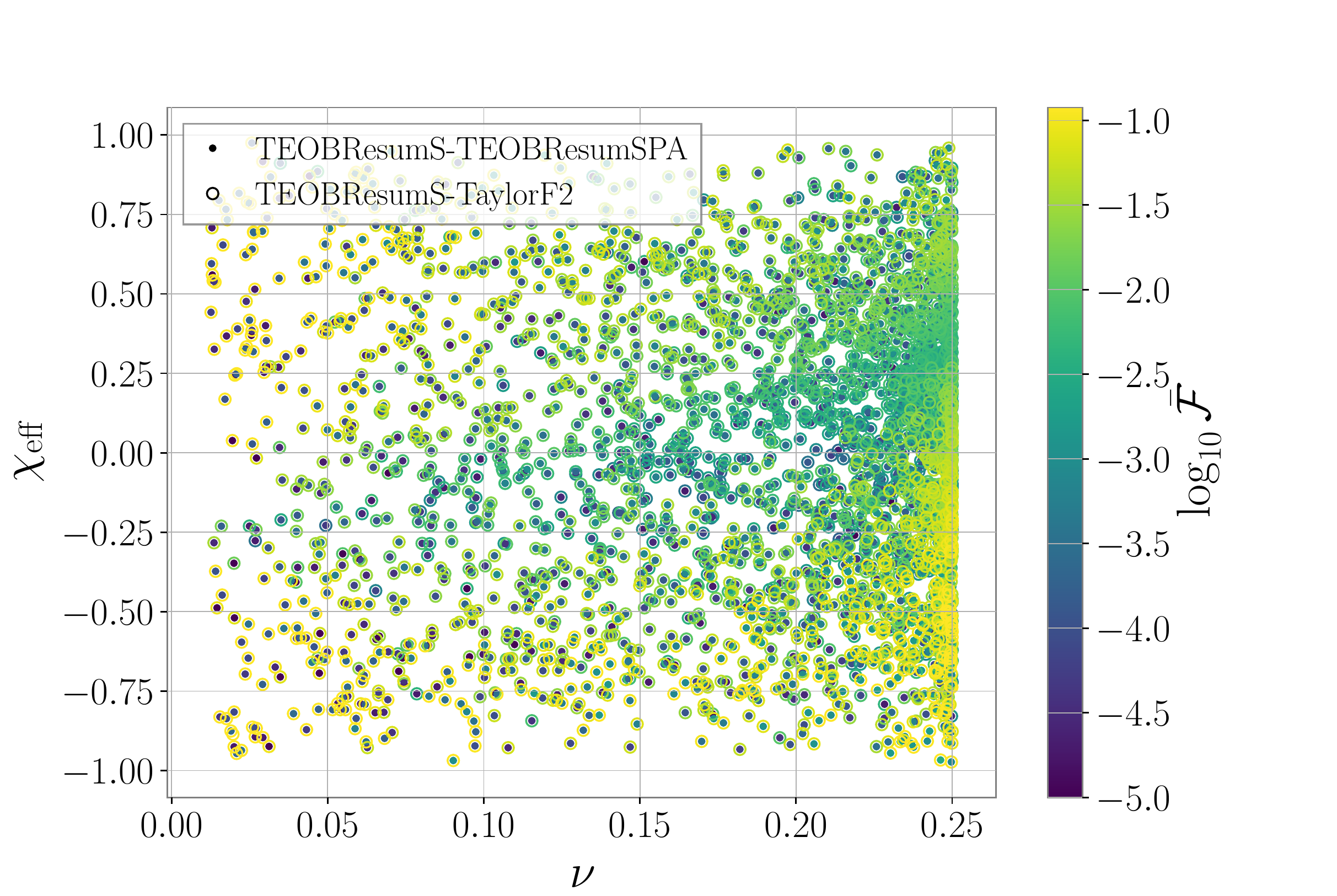}
  \caption{Mismatches $\bar{\cal F}$ between TD \TEOB{} and
   \TEOB{PA} or  the 3.5PN-accurate {\tt TaylorF2} on the frequency range $[0.02,1]$~Hz 
   obtained via Eq.~\eqref{eq:barF} with the LISA noise curve~\cite{Cornish:2018dyw}.
   Top panel: for moderately large mass ratios ($\nu\leq 0.1$, corresponding to $q\gtrsim 7.8$) 
   and massive ($M > 10^4\Mo$) BBH systems, mismatches between \TEOB{} 
  and {\tt TaylorF2} become large, reaching $\bar{\mathcal{F}} \sim 0.1$.
  Bottom panel: {\tt TaylorF2} is effectual only for moderate values of the effective spin
  parameter $\chi_{\rm eff}\equiv S_1/(m_1 M)+S_2/(m_2 M)$.}
  \label{fig:fbar_lisa}
\end{figure}

The EOB SPA waveform can be used to compute 
accurate  inspiral waveforms of intermediate-mass 
BBH that will be observed 
by LISA in the mHz to Hz regime~\cite{Audley:2017drz} and 
possibly in the dHz regime by planned spaced-based 
detectors as DECIGO~\cite{Sato:2009zzb,Sedda:2019uro,Kawamura:2020pcg}.
Beside PN and EOB approximant, no other modeling technique is
available for this inspiral-to-late-inspiral regime~\footnote{We recall that 
Phenomenological approximants are based on fits to PN and EOB waveforms, 
and do not provide an independent description of this regime.}. 
We focus on LISA sources with total masses $\sim 10^3-10^5\Mo$ and mass
ratio up to $q \sim 80$
\footnote{Note that we do not simulate coalescences over the entire LISA frequency
band, but focus only on the last hours/days of inspiral. This is due to
the heavy computational burden represented by interpolating and working with  
waveforms evaluated on a frequency grid having $\Delta f \lesssim 1/{\rm year}$}.
Figure~\ref{fig:fbar_lisa} compares $\sim 10^4$ mismatches between 
\TEOB{}--\TEOB{PA} with \TEOB{}--{\tt TaylorF2} (at 3.5PN accuracy) 
computed with the LISA noise curve~\cite{Cornish:2018dyw}.
The top panel focuses on the $(\nu,M)$ dependence; the effect
of spin is described by $\chi_{\rm eff}\equiv S_1/(m_1M)+S_2/(m_2 M)$.
The figure highlights that {\tt TaylorF2} becomes more and more 
inaccurate when: (i) $M\gtrsim 10^{3}M_\odot$; 
(ii) when $q \gtrsim 8$ ($\nu\lesssim 0.0988$) and 
(iii) when spin magnitudes are not moderate. Specifically, 
the worst {$\TEOB{}$ -- \tt TaylorF2} mismatch ($\bar\F = 0.397$) 
corresponds to $(M/M_\odot, q, \chi_1, \chi_2) = (7670, 55, -0.84, -0.62)$,
and detection losses up to $78\%$.
In addition, we use again Eq.~\eqref{eq:threshold}, and consider SNR
of 20 and 
100~\cite{Audley:2017drz,Cutler:2019krq} to find threshold
mismatches of $\bar{\F}_{20} = 5\times 10^{-3}, \bar{\F}_{100} = 2
\times 10^{-4}$ for $D=4$. \TEOB{}-\TEOB{PA}
mismatches always lie below the lower
threshold $\F_{20}$, while $\approx 43\%$ of the simulated
signals also satisfy the stricter requirement with
$\bar{\F}_{100}$. On the contrary, 
$59\%$ of the PN waveforms are not faithful at SNR 20, and $42\%$ of
them correspond to systems with $q > 8$. Therefore, {\tt TaylorF2}
is not a robust choice for parameter estimation of these BBH sources. 
The discrepancy between PN and EOB further increases
considering higher frequencies; 
the analysis of intermediate-mass BBH in the the dHz regime (e.g. DECIGO band)
will require fast and accurate FD EOB models like \TEOB{PA} (see Supplementary Material).

In summary, the EOB SPA method proposed here will crucially support 
both ground-based and spaced-based observations of long
(minutes-to-years) GW transient from compact binaries, whose 
analysis is a formidable challenge for the years to come.
\\

\acknowledgments
This work was funded by the Deutsche Forschungsgemeinschaft (DFG) under Grant No. 406116891 within the Research Training Group RTG 2522/1.
S.~B. acknowledge support by the EU H2020 under ERC Starting Grant, no.~BinGraSp-714626.  
Data analysis was performed on the supercomputer ARA at Jena. We acknowledge the computational resources provided
by Friedrich Schiller University Jena, supported in part by DFG grants
INST 275/334-1 FUGG and INST 275/363-1 FUGG.
Data postprocessing was performed on the Virgo ``Tullio'' server 
in Torino, supported by INFN.
{\tt TEOBResumS} and {\tt TEOBResumSPA} are both publicly available at

\url{https://bitbucket.org/eob_ihes/teobresums/}

\noindent
This research has made use of data, software and/or web tools obtained 
from the Gravitational Wave Open Science Center (\url{https://www.gw-openscience.org}), 
a service of LIGO Laboratory, the LIGO Scientific Collaboration and the 
Virgo Collaboration. LIGO is funded by the U.S. National Science Foundation. 
Virgo is funded by the French Centre National de Recherche Scientifique (CNRS), 
the Italian Istituto Nazionale della Fisica Nucleare (INFN) and the 
Dutch Nikhef, with contributions by Polish and Hungarian institutes.

\if\IncludeSM1

\clearpage

\widetext

\setcounter{equation}{0}
\setcounter{figure}{0}
\setcounter{table}{0}
\setcounter{page}{1}
\makeatletter
\renewcommand{\theequation}{S\arabic{equation}}
\renewcommand{\thefigure}{S\arabic{figure}}
\renewcommand{\bibnumfmt}[1]{[S#1]}

\begin{center}
\textbf{\large Supplemental Material}
\end{center}
\ifx\IncludeSM\undefined
\def\IncludeSM{0}
\fi

\if\IncludeSM0 
\documentclass[notitlepage,aps,prl,showpacs,superscriptaddress,groupedaddress,nofootinbib,floatfix]{revtex4-1}
\usepackage{graphicx,psfrag}
\usepackage{mathrsfs}
\usepackage{amsmath,amsfonts,amssymb}
\usepackage{multirow}
\usepackage{comment}
\usepackage{ulem}
\usepackage{hyperref}
\usepackage{enumitem}
\usepackage{morefloats}
\usepackage{bm} 
\newcommand{\be}{\begin{equation}}
\newcommand{\ee}{\end{equation}}
\newcommand{\bel}{\begin{align}}
\newcommand{\eel}{\end{align}}
\def\lm{\ell m}
\def\p{\partial}
\def\non{\nonumber}                     
\def\e{{\rm e}}
\def\i{{\rm i}}
\def\f{{\rm f,~BH}}
\def\gccm{{\rm g\,cm^{-3}}}
\def\Msun{{M_{\odot}}}
\def\Mpc{{\rm Mpc}}
\def\GMc2{{\rm G M_{\odot} c^{-2}}}
\def\Mpc{{\rm Mpc}}
\def\eps{\epsilon}
\def\eps{\epsilon}
\def\B{\mathcal{B}}
\def\I{\mathcal{I}}
\def\M{\mathcal{M}}
\def\O{\mathcal{O}}
\def\R{\mathcal{R}}
\def\vareps{\varepsilon}
\def\vrho{\varrho}
\def\check{$\checkmark$}
\def\cross{$\times$}
\def\l{\ell}
\def\lm{{\ell m}}
\def\hE{\hat{E}}
\def\mns{M_\text{NS}}
\def\Rns{R_\text{NS}}
\def\kt2{\kappa^\text{T}_2}
\def\Mmax{M_\text{TOV}^\text{max}}
\def\Mo{{\rm M_{\odot}}}
\def\kt2{\kappa^\text{T}_2}
\def\Mmax{M_\text{max}}
\def\Rmax{R_\text{max}}
\def\Rmax{R_\text{max}}
\def\Mmaxll{M_\text{max}^\text{LB}}
\def\params{{\boldsymbol{\theta}}}
\def\F{\mathcal{F}}
\def\Fbar{\bar{\mathcal{F}}}
\def\MM{\Fbar}
\newcommand{\nrpm}{\texttt{NRPM}}
\newcommand{\TEOB}[1]{\texttt{TEOBResumS{#1}}}
\usepackage{color}
\definecolor{cyan}{rgb}{0,0.9,0.9}
\definecolor{orange}{rgb}{0.9,0.5,0}
\definecolor{magenta}{rgb}{1,0,1}
\definecolor{purple}{rgb}{0.8,0.4,0.8}
\definecolor{gray}{rgb}{0.5,0.5,0.5}
\newcommand{\mb}[1]{{\textcolor{blue}{\texttt{MB: #1}} }}
\newcommand{\bs}[1]{{\textcolor{green}{\texttt{SB: #1}} }}
\newcommand{\rg}[1]{{\textcolor{orange}{\texttt{RG: #1}} }}
\newcommand{\an}[1]{{\textcolor{red}{\texttt{AN: #1}} }}

\newcommand{\todo}[1]{{\textcolor{red}{TODO: [#1]}}} 
\newcommand{\red}[1]{{\textcolor{red}{#1}}} 
\newcommand{\newtxt}[1]{{\textcolor{red}{#1}}} 
\newcommand{\oldtxt}[1]{{\textcolor{gray}{\sout{#1}}}} 
\newcommand{\oldnewtxt}[2]{{\textcolor{gray}{\sout{#1}}}\red{#2}} 
\newcommand{\timesto}[1]{\times 10^{#1}}
\begin{document}
\title{Supplemental Material:\\ Fast, faithful, frequency-domain effective-one-body waveforms\\ for binary neutron star inspiral-mergers}
\author{Rossella \surname{Gamba}${}^{1}$}
\author{Sebastiano \surname{Bernuzzi}${}^{1}$}
\author{Alessandro \surname{Nagar}${}^{2,3,4}$}
\affiliation{${}^1$Theoretisch-Physikalisches Institut, Friedrich-Schiller-Universit{\"a}t Jena, 07743, Jena, Germany}  
\affiliation{${}^2$Centro Fermi - Museo Storico della Fisica e Centro Studi e Ricerche Enrico Fermi, Rome, Italy}  
\affiliation{${}^3$INFN Sezione di Torino, Via P. Giuria 1, 10125 Torino, Italy}  
\affiliation{${}^4$Institut des Hautes Etudes Scientifiques, 91440 Bures-sur-Yvette, France}
\date{\today}
\maketitle
\fi 


\section{Stationary Phase Approximation (SPA)}
\label{AppA:SPA}

We review here the SPA approach to compute the Fourier
transform of a signal in the time domain (TD).
Given a real TD function $h(t)$, defined as 
\begin{equation}
\label{wf}
h(t)=2 a(t) \cos \phi(t)=a(t) e^{-i \phi(t)}+a(t) e^{i \phi(t)} \ ,
\end{equation}
with $\dot{\phi}(t) \equiv 2 \pi F(t)>0$, its Fourier transform 
$\tilde{h}$ can be expressed as
\be
  \tilde{h}(f) =\tilde{h}_{m}(f)+\tilde{h}_{p}(f)  \ ,
\ee
where
\begin{align}
  \tilde{h}_{m}(f) & \equiv \int_{-\infty}^{\infty} d t\, a(t) e^{i(2 \pi f t-\phi(t))}  \ , \\ 
  \tilde{h}_{p}(f) & \equiv \int_{-\infty}^{\infty} d t \, a(t) e^{i(2 \pi f t+\phi(t))} \ .
\end{align} 
Since the integrands oscillate rapidly, the largest contribution to
the integral comes from the vicinity of the stationary points of their 
phase (if such points exist).  Assuming $f > 0$, only the phase 
of $\tilde{h}_{m}$ is stationary and we can therefore neglect the 
contribution of $\tilde{h}_{p}$ to obtain
\be
\label{htilde}
\tilde{h}(f) \simeq \tilde{h}_{m}(f) \simeq
\int_{-\infty}^{\infty} d t a(t) e^{i \psi_{f}(t)} \ ,
\ee
with
\be
\psi_{f}(t) \equiv 2 \pi f t-\phi(t) \ .
\ee
Defining the saddle point of $ \psi_{f}(t)$ as $t_f$, we find that the largest contribution to the integral is given by $F(t_f) = f$, i.e by points 
in the Fourier space in which $f$ equals the instantaneous GW frequency $F$. When the second derivative of the phase is non vanishing, 
one can then estimate Eq.~\eqref{htilde} as
\begin{align}
  \label{psi_approx}
  \psi_{f}(t) & \simeq \psi_{f}\left(t_{f}\right)-\pi
  \dot{F}\left(t_{f}\right)\left(t-t_{f}\right)^{2} \ ,\\
  a(t) & \simeq a\left(t_{f}\right) \ .
\end{align}
The Gaussian integral one obtains by plugging Eq.~\eqref{psi_approx} into Eq.~\eqref{htilde} can then be easily solved to obtain
\be 
\label{uspa}
\tilde{h}^{\mathrm{SPA}}(f)=\frac{a\left(t_{f}\right)}{\sqrt{\dot{F}\left(t_{f}\right)}} e^{i\left[\psi_{f}\left(t_{f}\right)-\pi / 4\right]}\ .
\ee
The above expression can be applied straightforwardly to each mode
$h_{\ell m}$. The modes are then be combined to obtain the two 
polarizations as  
\begin{subequations}
\begin{align}
\label{eq:spa_hphc}
\tilde{h}_{+}(f)&=\frac{1}{2} \sum_{\ell\geq 2}\sum_{ m>0}^{\ell} \tilde{A}_{\ell m}^{\rm SPA} e^{i \Psi_{\ell m}^{\rm SPA}}\left[_{-2}Y_{\ell m}^{*}+(-)^{\ell}{}_{-2} Y_{\ell-m}\right] \ ,\\
\tilde{h}_{\times}(f)&= \frac{i}{2} \sum_{\ell\geq 2}\sum_{m>0}^{\ell} \tilde{A}_{\ell m}^{\rm SPA} e^{i \Psi_{\ell m}^{\rm SPA}} \left[_{-2}Y_{\ell m}^{*}+(-)^{\ell+1}{}_{-2} Y_{\ell-m} \right] \ .
\end{align}
\end{subequations}

The SPA method is applied to the TD waveform generated by \TEOB{}, a state-of-the-art effective EOB approximant for 
spin-aligned binaries~\cite{Damour:2014sva,Bernuzzi:2015rla,Nagar:2017jdw,Nagar:2018zoe,Nagar:2018plt,Akcay:2018yyh,Nagar:2019wds,Nagar:2020pcj}.
Although the procedure is straightforward, we comment on two technical details.
First, the SPA implementation requires the evaluation of (second) time derivatives
of the phase of each multipole, $\ddot{\phi}_{\ell m}(t)$. 
In order to employ the SPA together with the post-adiabatic
approximation of the dynamics~\cite{Nagar:2018gnk}, we thus implement 
a fourth-order finite difference formula for non-uniform grids derived in
the standard way using Lagrangian interpolants. A further speed up 
of the waveform generation can in principle be achieved by substituting 
the numerical derivatives $\dot\phi_{\ell m}, \ddot\phi_{\ell m}$ 
with their explicit expressions in terms of the EOB dynamical variables. 
Extensive tests shows that the numerical differentiation is satisfactory 
(at least for the applications presented in the main paper), and therefore we leave such 
an improvement to future work.

\begin{figure*}[t]
  \centering 
  \includegraphics[width=.45\textwidth]{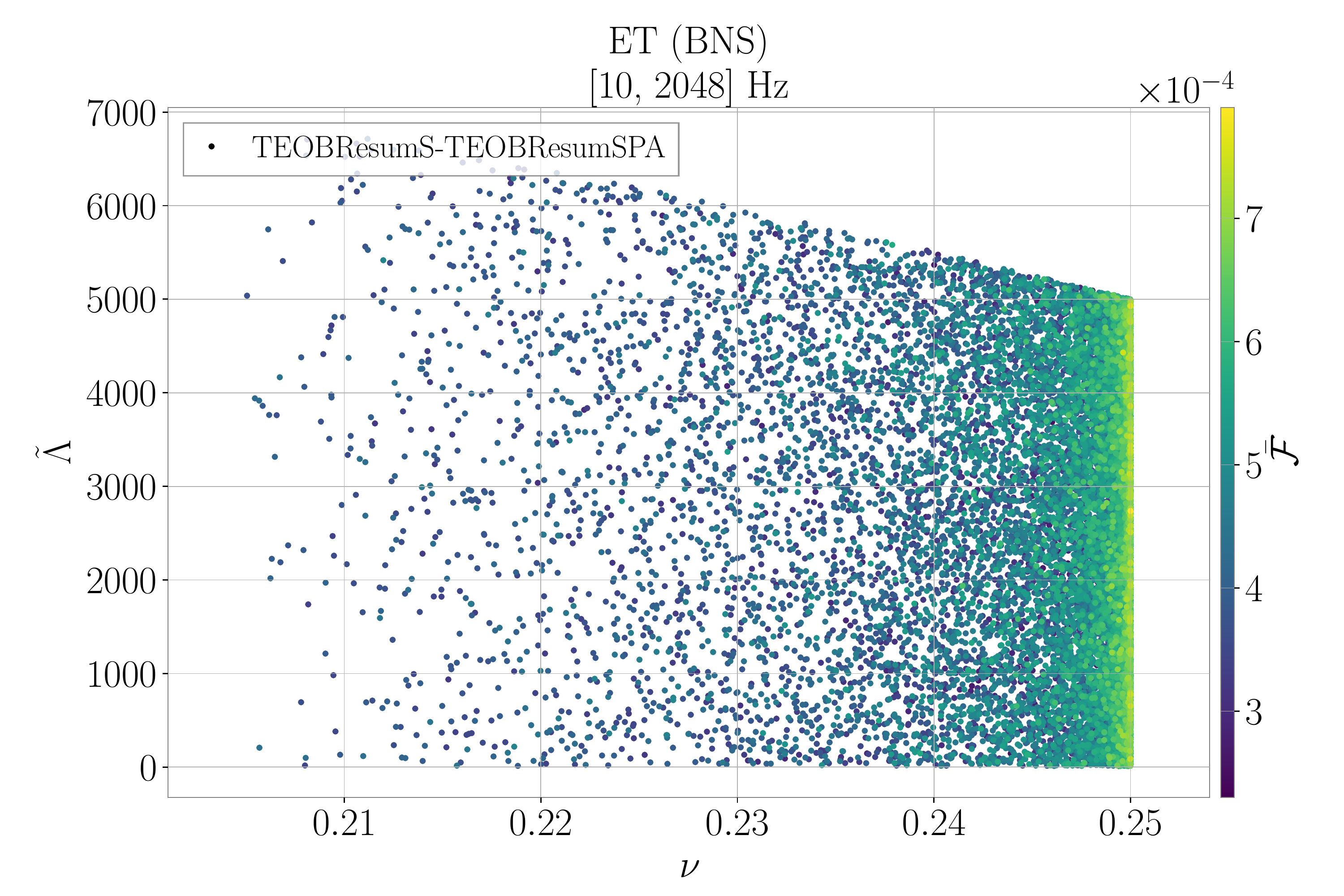}
  \includegraphics[width=.45\textwidth]{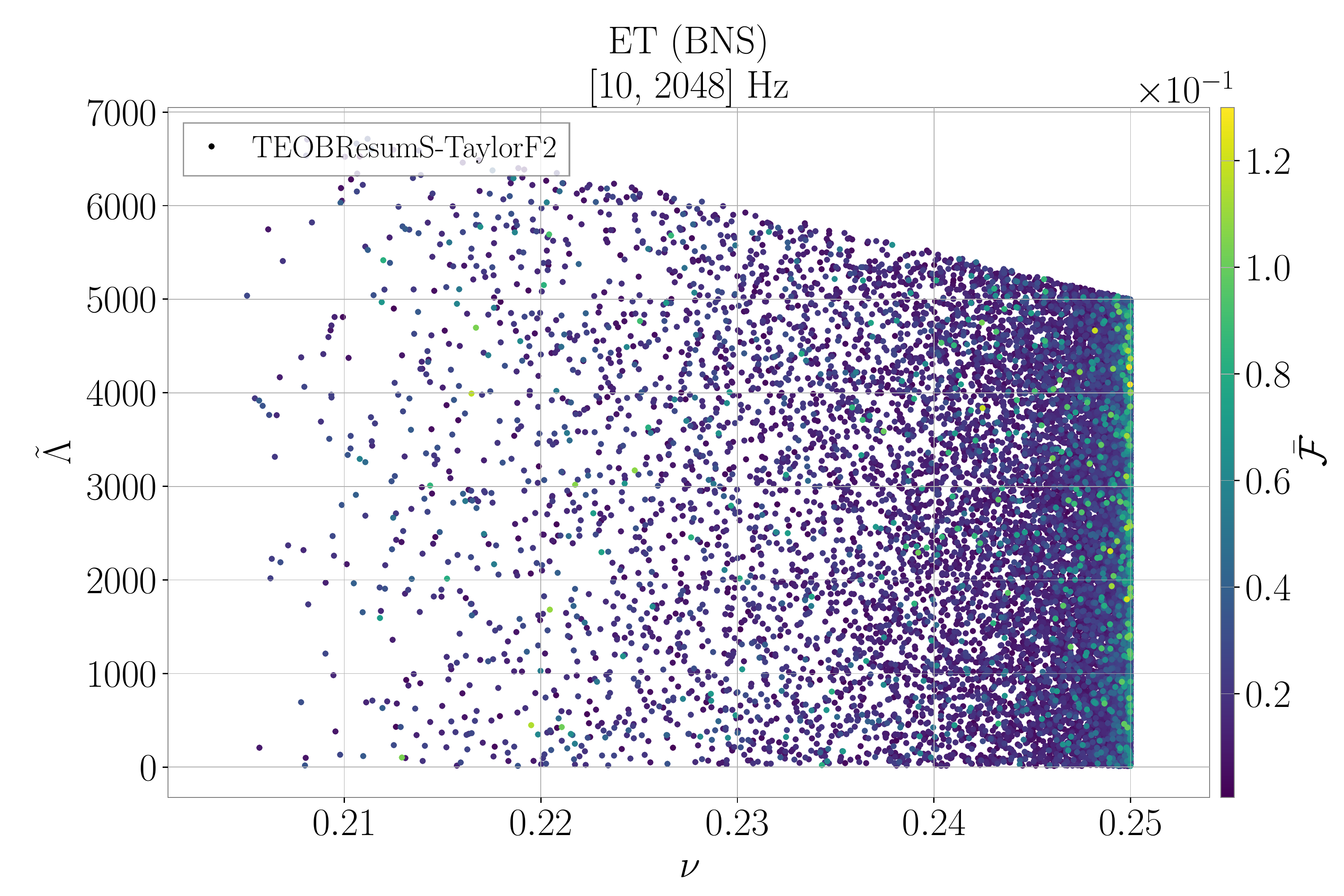}
  \includegraphics[width=.45\textwidth]{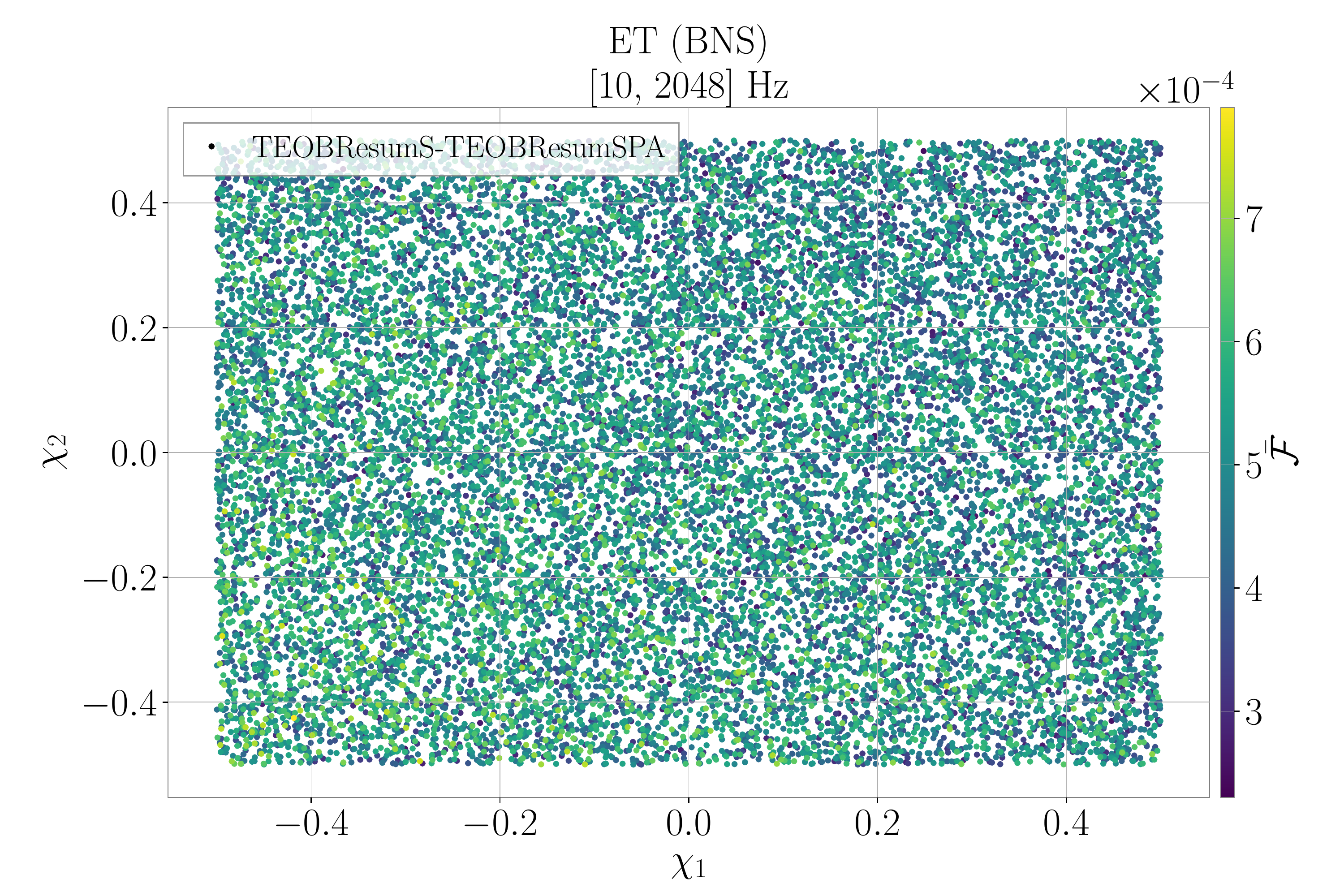}
  \includegraphics[width=.45\textwidth]{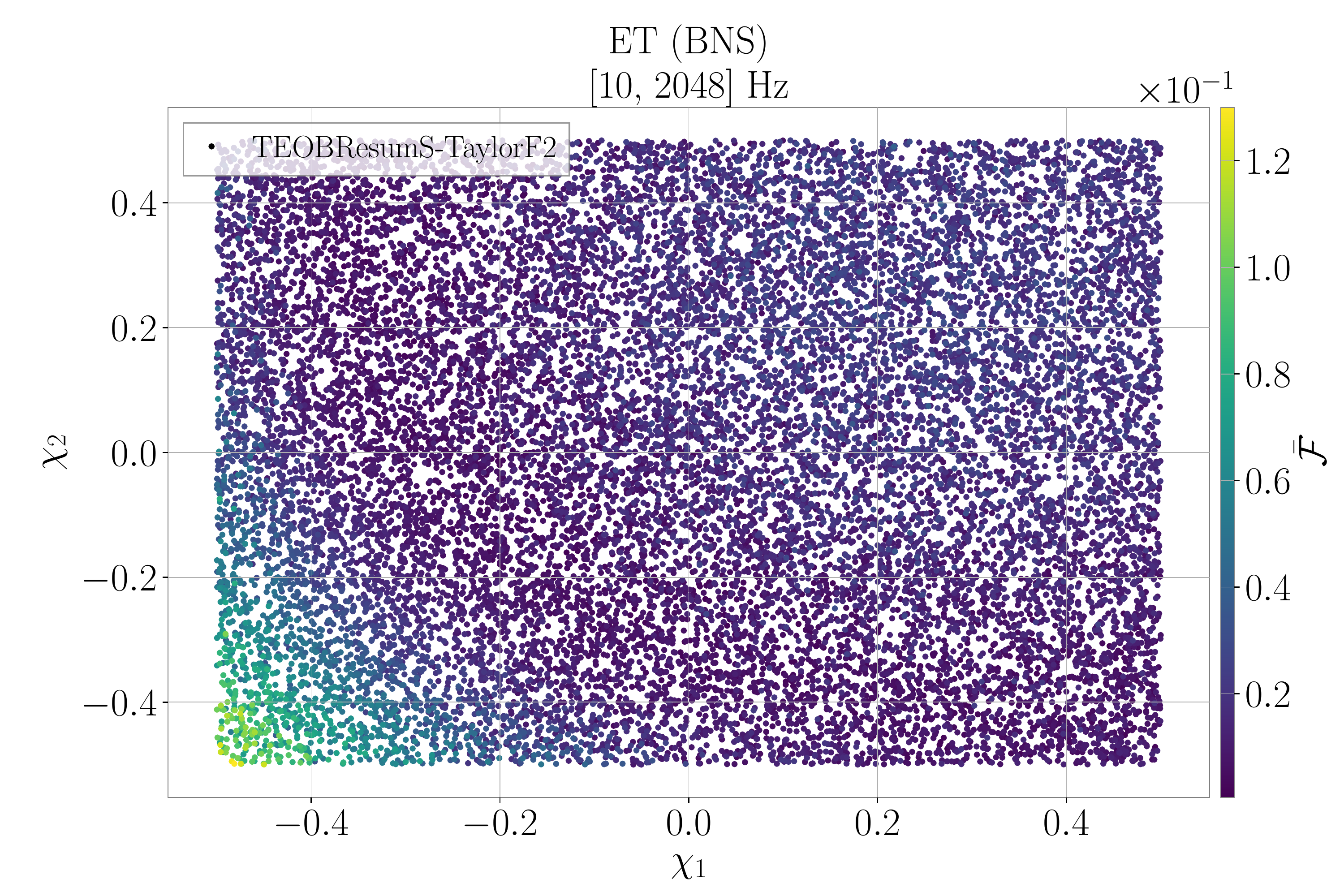}
  \caption{Mismatches computed between $\TEOB{}$ time-domain waveforms and $\TEOB{PA}$
  or {\tt TaylorF2} for BNS systems between [10, 2048] Hz with the ET noise curve~\cite{Evans:2016mbw}. 
  While $\TEOB{PA}$ always displays mismatches better than $\sim 10^{-4}$, {\tt TaylorF2} reaches a mismatch of $12\%$ 
  for large misaligned spins (bottom right panel).}
  \label{fig:sup_fbar_et}
\end{figure*}

\begin{figure*}[t]
  \centering 
  \includegraphics[width=.45\textwidth]{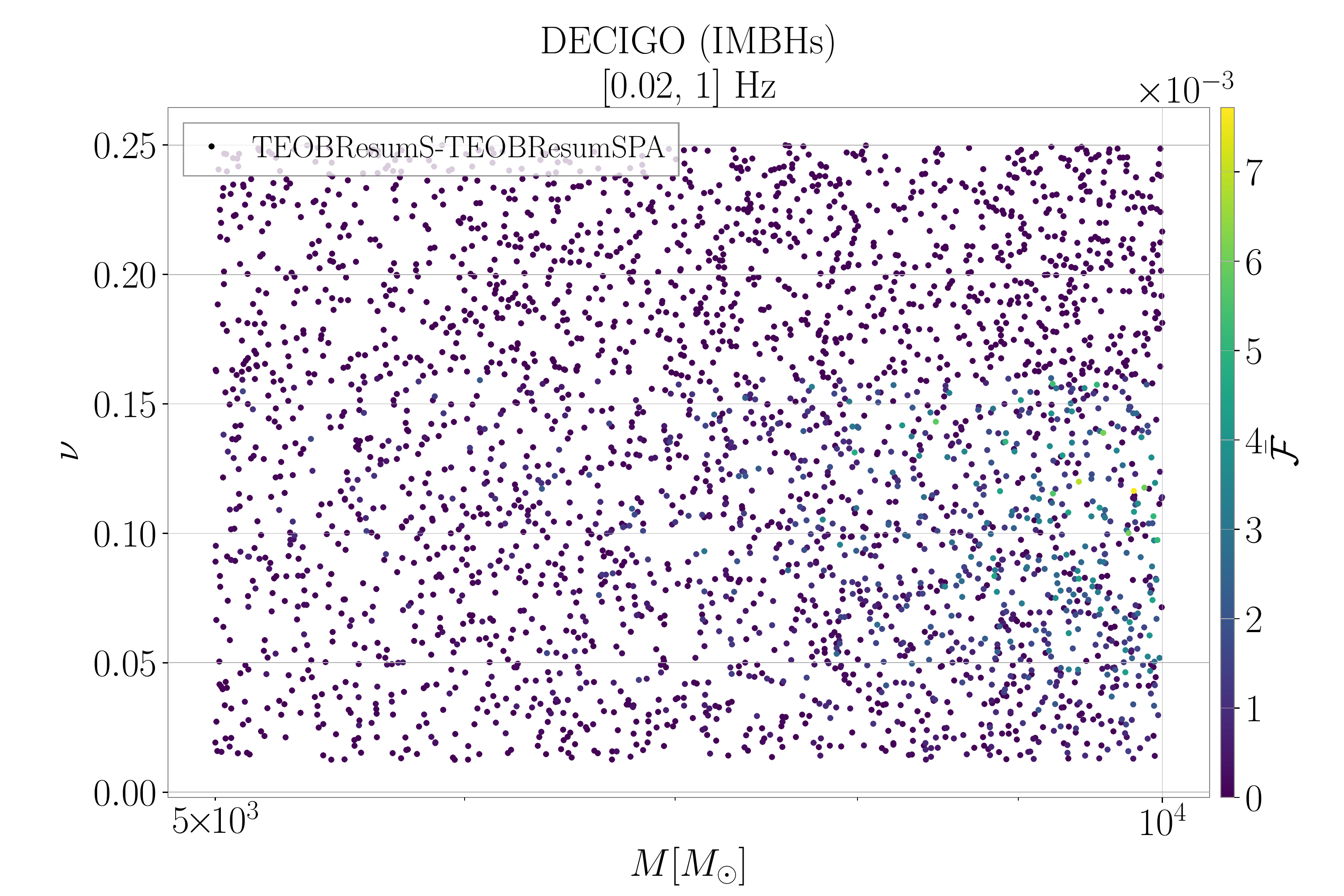}
  \includegraphics[width=.45\textwidth]{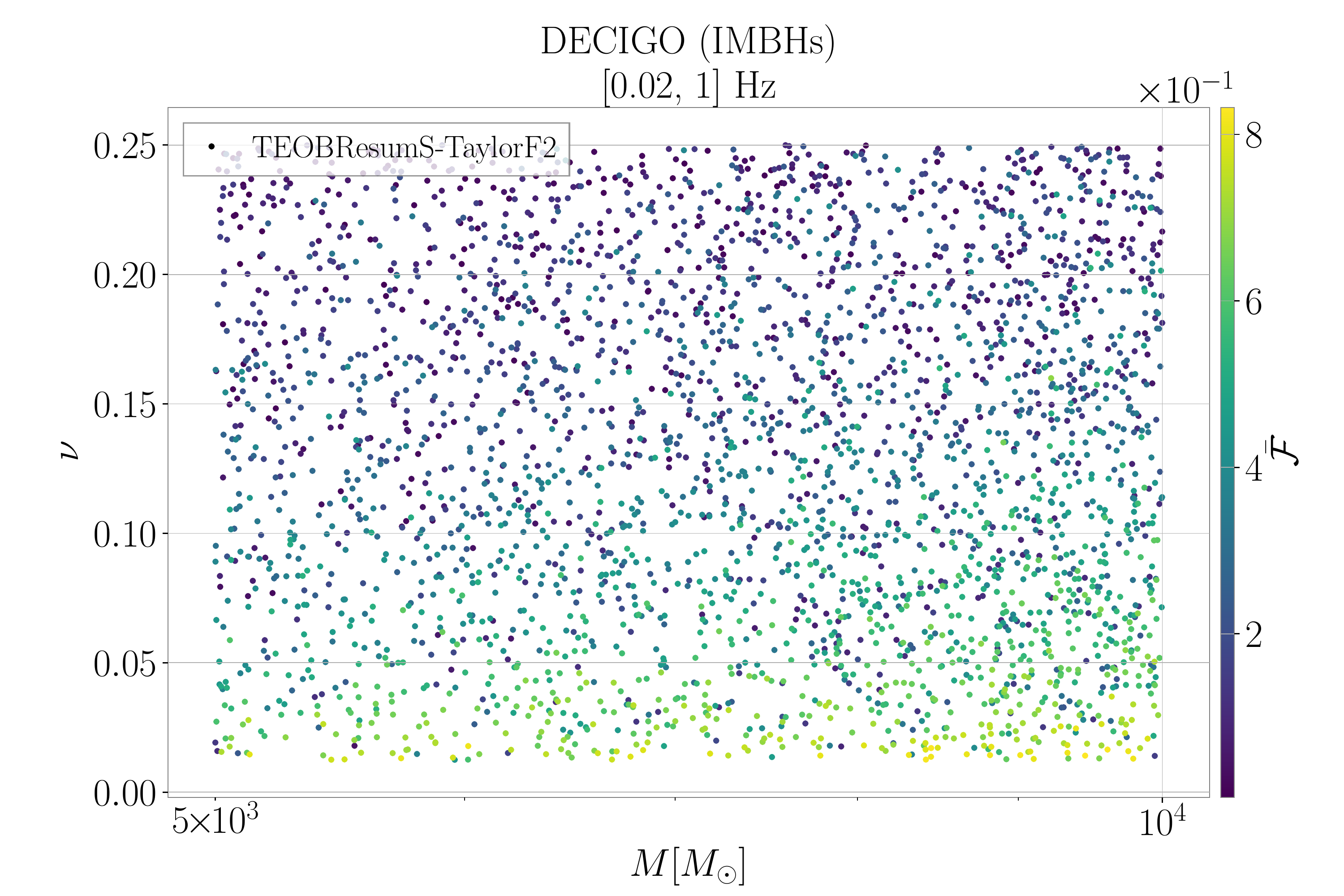}
  \includegraphics[width=.45\textwidth]{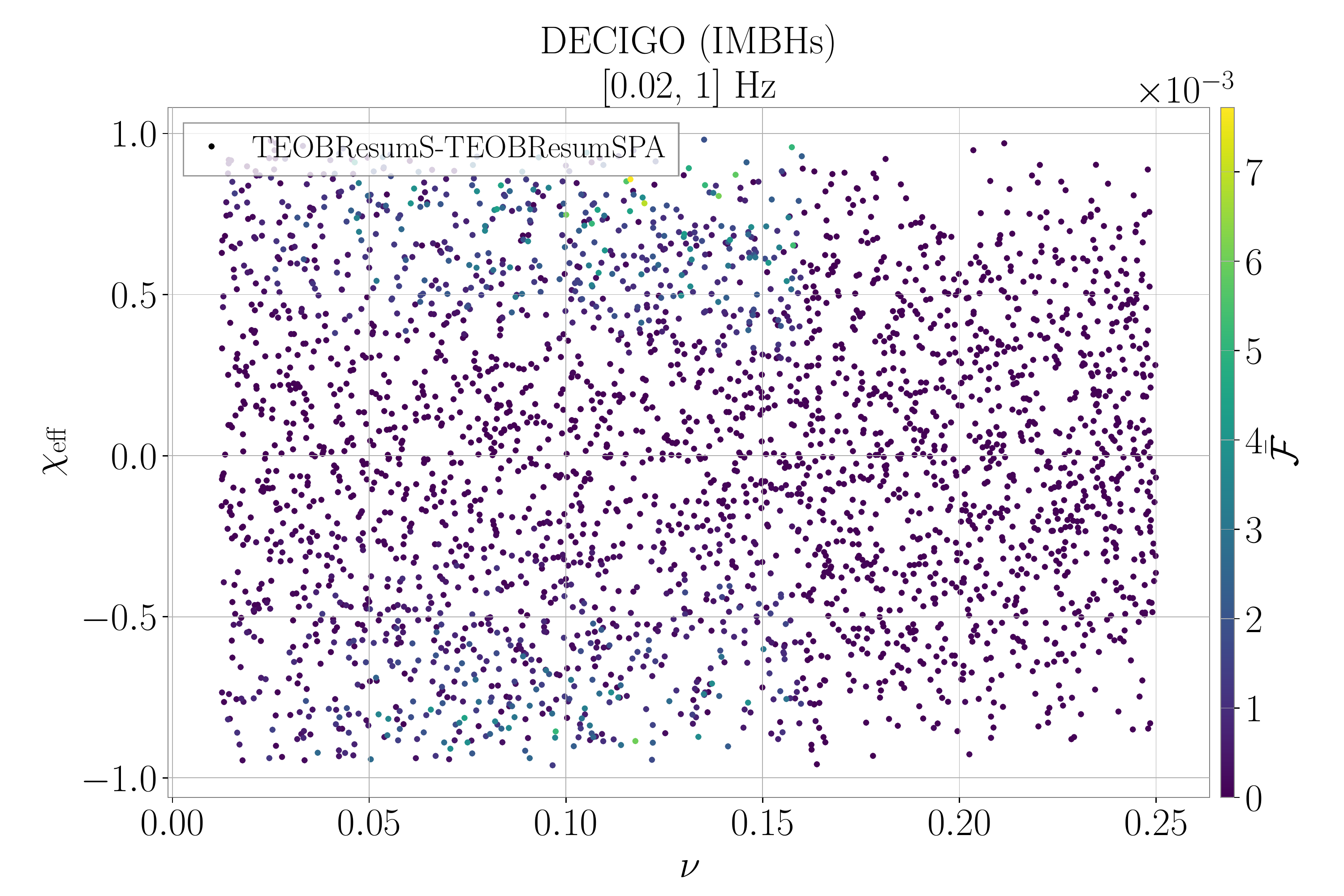}
  \includegraphics[width=.45\textwidth]{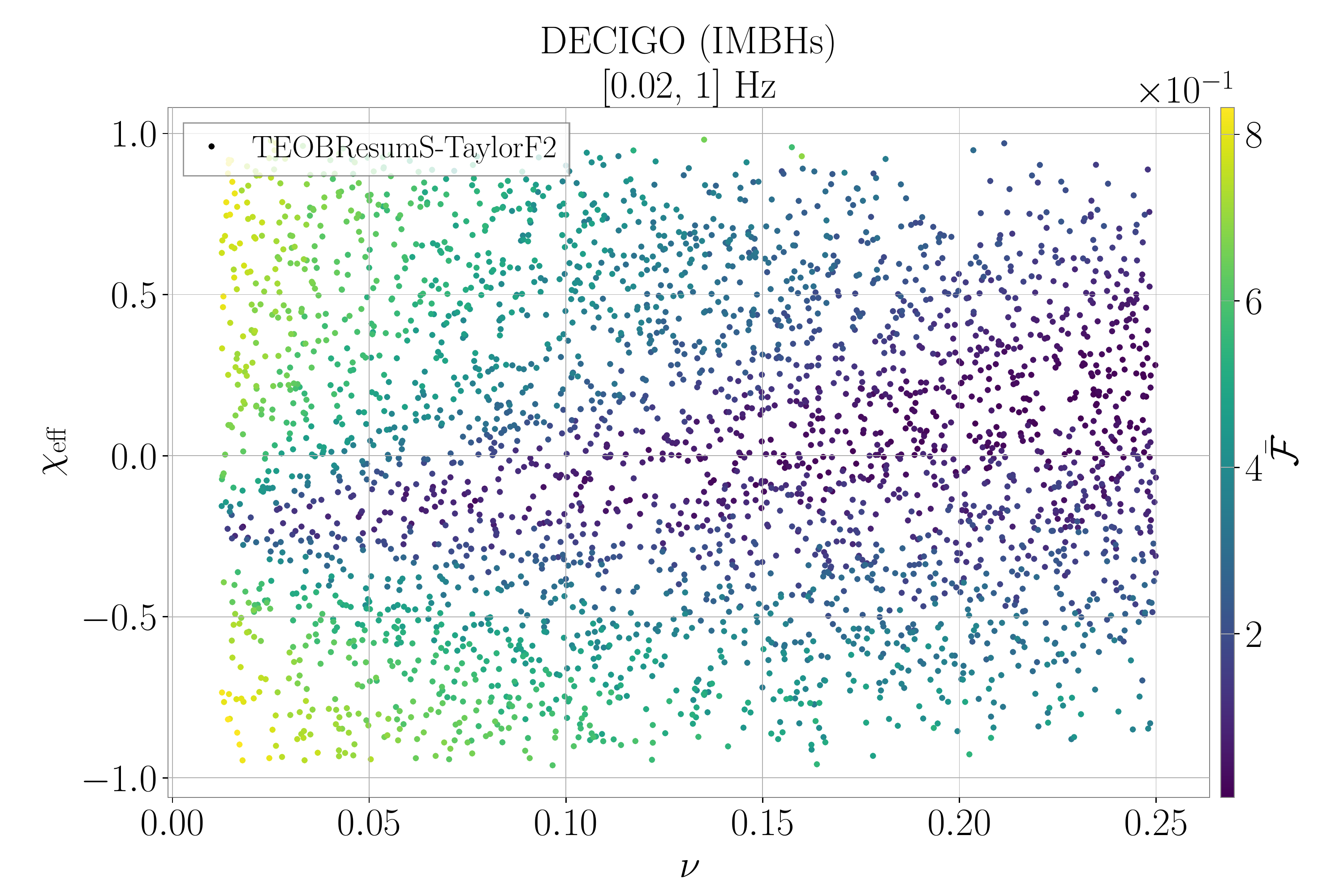}
  \caption{Mismatches computed between $\TEOB{}$ time-domain waveforms and $\TEOB{PA}$
  or 3.5PN accurate {\tt TaylorF2} for heavy IMBH systems in the frequency window
  $[0.02, 1]\,$Hz of the DECIGO noise curve~\cite{Sato:2009zzb}. 
  We use the effective spin parameter $\chi_{\rm eff}\equiv S_1/(m_1 M) + S_2/(m_2 M)$. 
  The large mismatches between $\TEOB{}$ and $\tt TaylorF2$ clearly demonstrate 
  that the latter is not even effectual for describing heavy asymmetric IMBH systems 
  in the DECIGO frequency band.}
  \label{fig:sup_fbar_decigo_heavy}
\end{figure*}

\begin{figure*}[t]
  \centering 
  \includegraphics[width=.45\textwidth]{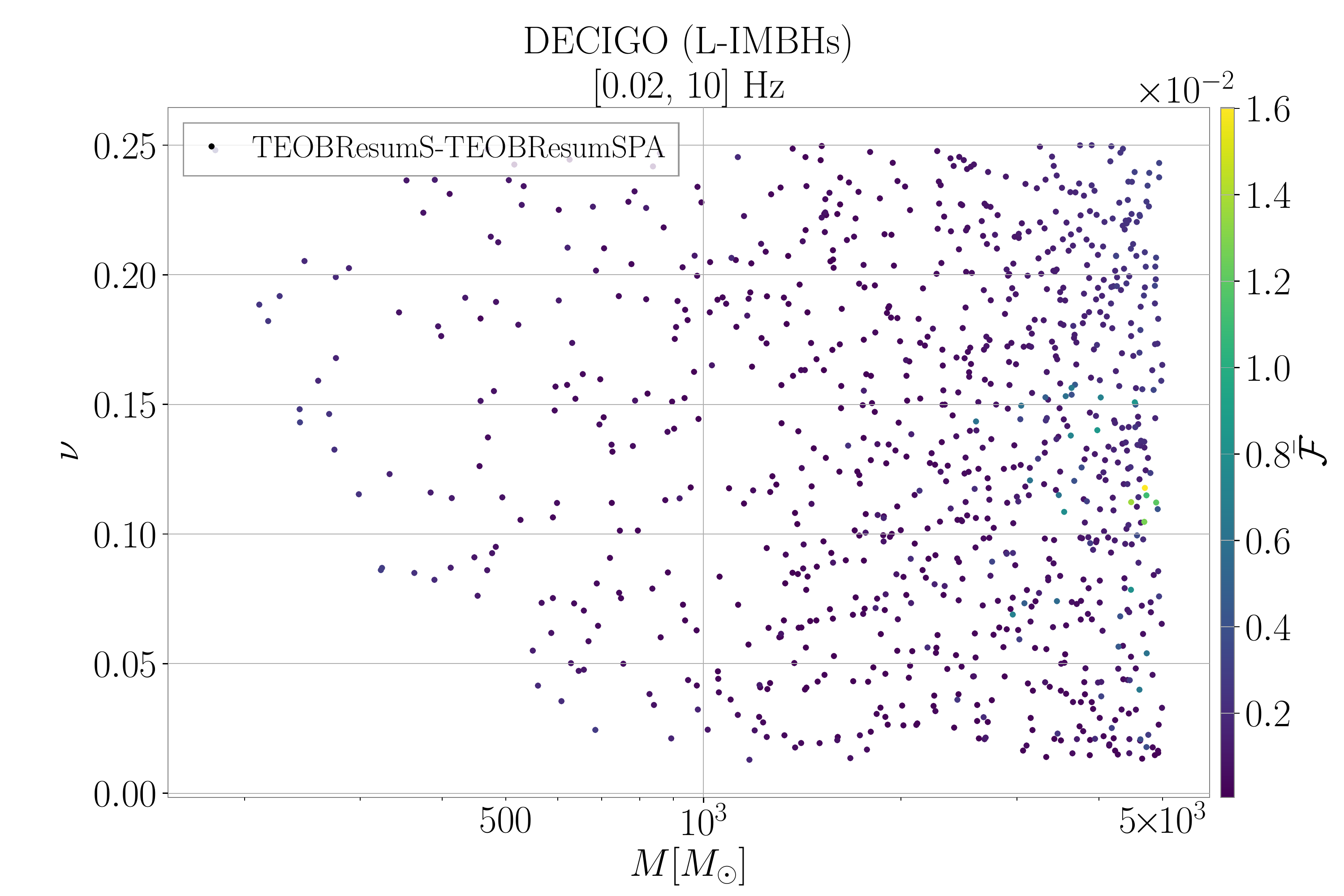}
  \includegraphics[width=.45\textwidth]{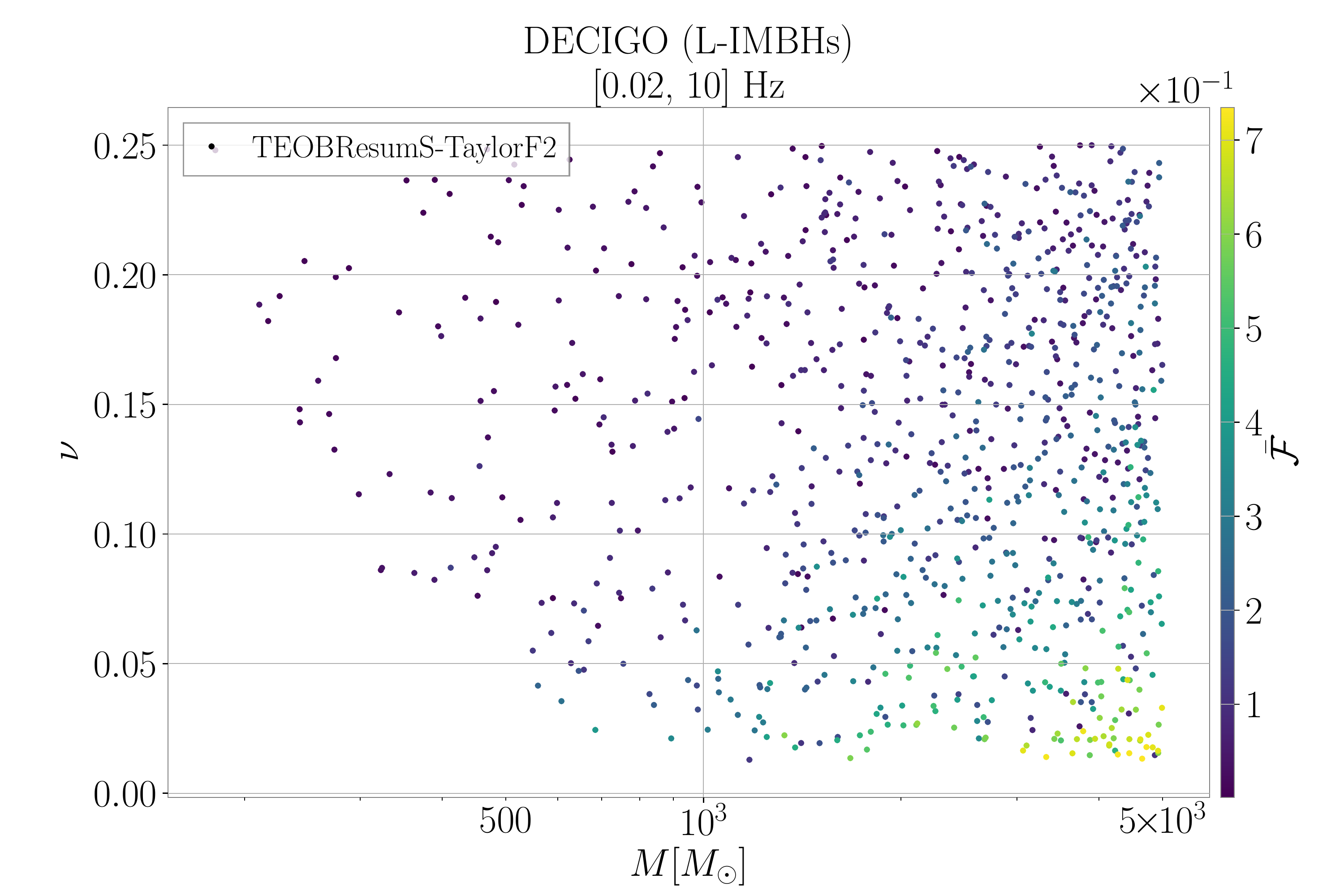}
  \includegraphics[width=.45\textwidth]{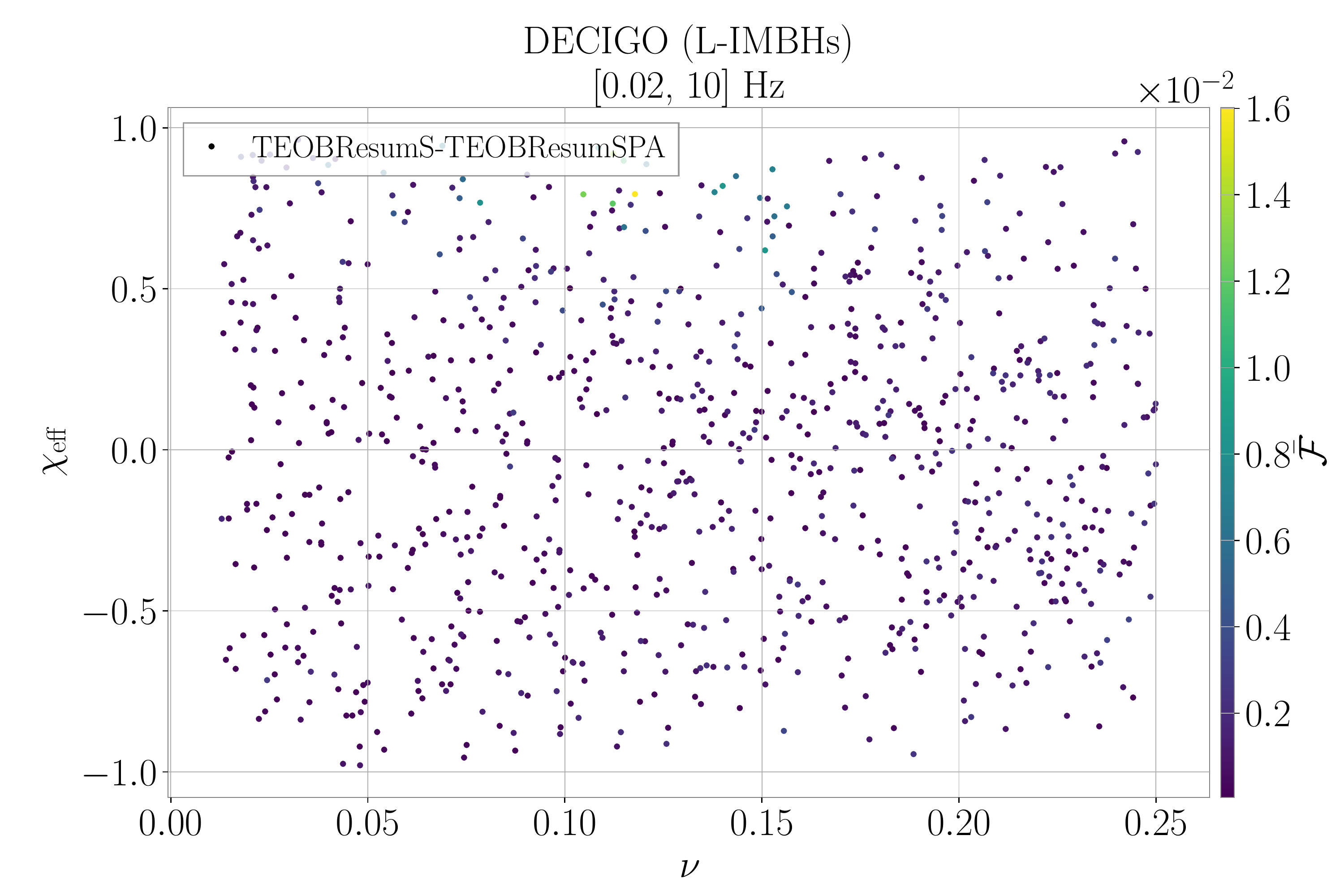}
  \includegraphics[width=.45\textwidth]{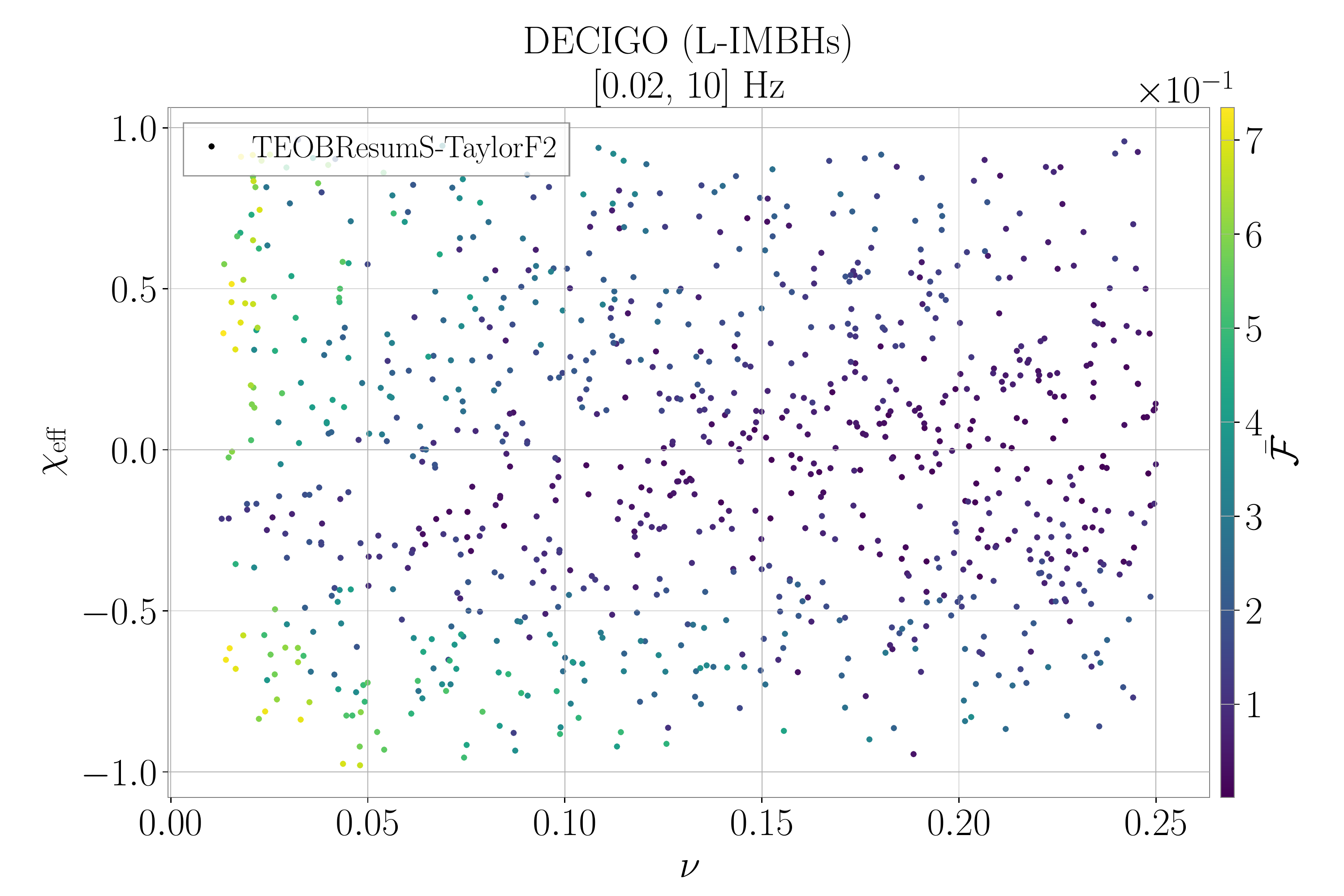}
  \caption{Mismatches computed between $\TEOB{}$ time-domain waveforms and $\TEOB{PA}$
  or 3.5PN-accurate {\tt TaylorF2} for light IMBH systems in the frequency window $[0.02, 10]\,$Hz of the DECIGO noise curve~\cite{Sato:2009zzb}. 
  We use the effective spin parameter $\chi_{\rm eff}\equiv S_1/(m_1 M) + S_2/(m_2 M)$. 
  The large mismatches between $\TEOB{}$ and $\tt TaylorF2$ clearly demonstrate 
  that the latter is not even effectual for describing light asymmetric IMBH systems 
  in the DECIGO frequency band.}
  \label{fig:sup_fbar_decigo_light}
\end{figure*}
In general, a direct numerical SPA of the time-windowed modes does not
allow for the computation of the FD waveform over the entire frequency
range of interest, which from the initial frequency $f_0$  
reaches up to the Nyquist frequency. 
Moreover, at a given time $t=\bar{t}$, the
instantaneous frequency of each mode $f_{\ell m}\equiv \dot{\phi}_{\ell m}(\bar{t})/(2\pi)$ 
is different for different modes. 
Therefore, in order to correctly compute the waveform
polarizations, it is necessary to: (i) identify a threshold value or frequency 
$f^{\rm max}_{\ell m}$ up to which the SPA maintains its validity and provide 
a continuation to $f >f^{\rm max}_{\ell m}$, and (ii) interpolate the various 
modes on a common frequency grid. 
For BNS systems, the continuation to frequencies higher than merger
can be obtained by completing the inspiral-merger model with 
a FD representation of the postmerger waveform, 
see e.g.~\cite{Breschi:2019srl}. Alternatively, for
inspiral-merger waveforms only, we implement 
a simple linear extrapolation of the FD phase and
a switch off the FD amplitude according to
\begin{align}
\Psi_{\ell m}(f > f^{\rm max}_{\ell m})& = \Psi_{\ell m}(f^{\rm max}_{\ell m}) + \Psi'_{\ell m}(f^{\rm max}_{\ell m})(f - f^{\rm max}_{\ell m})\ ,\\
\tilde{A}_{\ell m}(f > f^{\rm max}_{\ell m})  &=  \tilde{A}_{\ell m}(f^{\rm max}_{\ell m})\left(\frac{f^{\rm max}_{\ell m}}{f}\right)^{10/3}\ , 
\end{align}
where $f_{\ell m}^{\rm max} \equiv f_{\ell m}(t_{\rm max})$
and $t_{\rm max}$ is defined as the time when $\ddot{f}_{\ell m}=0$.
This choice of $t_{\rm max}$ and $f_{\ell m}^{\rm max}$ ensures that, 
when considering the $\ell=m=2$ mode, we also capture part of the frequency 
evolution {\it beyond} merger, as $f^{\rm mrg} < f_{22}^{\rm max}$. 
The interpolation on the common grid is performed after the extension to high frequencies on a 
uniformly spaced interval, whose $\Delta f$ is user-input. In GW parameter estimation, 
$\Delta f$ would be equal to the inverse of the length of the segment analyzed.

A C implementation of {\tt TEOBResumS} and {\tt TEOBResumSPA} with a
python interface is publicly available at
\begin{center}
  \url{https://bitbucket.org/eob_ihes/teobresums/}
\end{center}


\section{EOB-SPA faithfulness for BNS and IMBH systems in various frequency bands}
\label{AppA:Faith}

In this Section we collect additional mismatch studies to complement those
reported in the main text. We focus on: (i) BNS waveforms in the frequency
window of the Einstein Telescope (ET) \cite{Punturo:2010zz,
Sathyaprakash:2011bh, Maggiore:2019uih} (BNS, $f\in[10,2048]\,$Hz)
and (ii) Intermediate mass black holes (IMBHs) in the DECIGO~\cite{Sato:2009zzb, Kawamura:2020pcg}
frequency range. In particular, we consider
light IMBHs (L-IMBHs) with  $q\in [1, 80]$, $M/M_\odot \in[100, 5000]$,
and frequency range  $f\in[0.02,10]\,$Hz, and heavy IMBHs, 
with $M/M_\odot\in[5000, 10000]$ and frequency range $f\in[0.02,1]\,$Hz.

\subsection{BNS waveforms in the ET band}

Figure~\ref{fig:sup_fbar_et} displays the mismatches
between \TEOB{} and \TEOB{PA} (left panels) 
and \TEOB{} and the 3.5PN-accurate $\tt TaylorF2$ (right panels), computed 
using the theoretical ET \cite{Punturo:2010zz, Sathyaprakash:2011bh, Maggiore:2019uih}
noise curve~\cite{Evans:2016mbw}.  We consider BNS systems with
$m_{1,2}/M_\odot \in [1., 2.5]$, $\chi_{1,2} \in [-0.5, 0.5]$ and $\Lambda_{1,2} \in [10, 5000]$, 
and compute the mismatches from 10~Hz to 2048~Hz.
Noticeably, while $\TEOB{} - \TEOB{PA}$ mismatches are of $\O(10^{-4})$, $\TEOB - {\tt TaylorF2}$
differences can exceed $10\%$ for binaries with large negative spins. 
The largest mismatch ($\bar\F = 0.13$)  is obtained with a system 
of $(M/M_\odot, q, \chi_1, \chi_2, \tilde\Lambda) = (4.9, 1.02, -0.48, -0.5, 4089)$.
This value corresponds to a threshold SNR of $5$, which lies below the 
detection threshold SNR, nominally  taken to be $\rm SNR_{\rm det} = 8$. 
In general {\tt TaylorF2} is found to get progressively inaccurate as
the magnitude of the anti-aligned spins is increased.
This implies that the use of the standard 3.5PN-accurate {\tt TaylorF2} for GW 
searches can lead to detection losses up to $\sim 34\%$ when considering 
very extreme BNS systems.

\subsection{IMBHs waveforms in the DECIGO band}
Let us finally turn to considering the DECIGO detector~\cite{Sato:2009zzb, Kawamura:2020pcg}. 
We compute mismatches for heavy IMBHs, having $q\in[1,80]$ and $M/\Msun \in [5000, 10000] $ 
in the frequency range $ f \in [0.02, 1]\,$Hz, and light IMBHs, having $M /\Msun \in [100, 5000]$ 
in the range $f \in [0.02, 10]$~Hz. The results are collected in Figs.~\ref{fig:sup_fbar_decigo_heavy} and~\ref{fig:sup_fbar_decigo_light}.
Unsurprisingly, the largest differences are found for heavy systems
with large mass ratios. The mismatches computed in the $[0.02, 1]\,$Hz
band are larger than the ones found with LISA, as DECIGO is expected
to be more sensitive in that regime. We find that $\TEOB{} - \TEOB{PA}$ 
mismatches are of $\O(10^{-4})$, with few exceptions reaching up to $\O(10^{-2})$.   
$\TEOB - {\tt TaylorF2}$ mismatches, instead, can grow as large as $80\%$ for binaries
with large spins and high mass ratio.
This value corresponds to potential detection losses of $99\%$ of the systems
having such properties.

\if\IncludeSM0 
\bibliography{refs,local}
\end{document}
\fi 

\fi

\end{document}